\documentclass{aa}
\usepackage{graphicx}
\usepackage{txfonts}
\usepackage[]{natbib}
\newcommand{\spitzer}{{\it Spitzer}}
\newcommand{\galex}{{\it GALEX}}
\newcommand{\mi}{$\mu$m}

\newcommand{\hii}{\ion{H}{ii}}
\newcommand{\hi}{\ion{H}{i}}
\newcommand{\mm}{M\,33}
\newcommand{\Ha}{H$\alpha$}
\newcommand{\msun}{M$_\odot$}

\defcitealias{2007A&A...476.1161V}{Paper I}
\defcitealias{2009A&A...493..453V}{Paper II}
\defcitealias{2009A&A...495..479C}{Paper III}
\begin{document}
   \title{Star formation in \mm: the radial and local relations with the gas}

   \author{S. Verley
          \inst{1,2}
          \and
          E. Corbelli
          \inst{1}
         \and
          C. Giovanardi
          \inst{1}
          \and
           L. K. Hunt
          \inst{1}
	}


   \institute{
Osservatorio Astrofisico di Arcetri - INAF, Largo E. Fermi 5, 50125
Firenze, Italy\\
\email{[edvige, giova, hunt]@arcetri.astro.it}
\and
Dept. de F\'\i sica Te\'orica y del Cosmos, Facultad de Ciencias, Universidad de Granada, Spain\\
\email{simon@ugr.es}
             }

   \date{Received; accepted}

  \abstract
   {}
   {In the Local Group spiral galaxy \mm, we investigate the correlation between the star 
   formation rate (SFR) surface density, $\Sigma_{\rm SFR}$, and the gas density $\Sigma_{\rm gas}$ (molecular, 
   atomic, and total). We also explore
whether there are other physical quantities, such as the hydrostatic pressure and dust
optical depth, which establish a good correlation with $\Sigma_{\rm SFR}$.}
   {We use  the \Ha, far-ultraviolet (FUV), and bolometric emission maps 
to infer the SFR locally at different spatial scales, and in radial bins
using azimuthally averaged values. Most of the local analysis is done using the highest spatial
resolution allowed by gas surveys, 180~pc. 
The Kennicutt-Schmidt (KS) law, $\Sigma_{\rm SFR} \propto \Sigma_{\rm gas}^n$ is analyzed by three 
statistical methods.} 
   {At all spatial scales, with \Ha\ emission as a SFR tracer,
   the KS indices $n$ are always steeper than those derived with  
   the FUV and bolometric emissions. We attribute this to the lack of \Ha\ emission in low
   luminosity regions where most stars form in small clusters with an incomplete 
   initial mass function at their high mass end.
   For azimuthally averaged values the depletion timescale for the molecular gas is constant, 
   and the KS index is $n_{\rm H_2}=1.1\pm 0.1$. 
   Locally, at a spatial resolution of 180~pc, the correlation between $\Sigma_{\rm SFR}$
   and $\Sigma_{\rm gas}$ is generally poor, even though it is tighter with the molecular and total gas
   than with the atomic gas alone. Considering only
   positions where the CO~$J=1-0$ line is above the 2-$\sigma$ detection threshold and taking into 
   account uncertainties in $\Sigma_{\rm H_2}$ and $\Sigma_{\rm SFR}$, 
   we obtain a steeper KS index than obtained with radial averages: 
   $n_{\rm H_2}=2.22\pm 0.07$ (for FUV and bolometric SFR tracers), flatter than that
   relative to the total gas ($n_{\rm H_{tot}}=2.59\pm 0.05$). The gas depletion timescale is therefore 
   larger in regions of lower $\Sigma_{\rm SFR}$. Lower KS indices 
   ($n_{\rm H_2}=1.46\pm 0.34$ and $n_{\rm H_2}=1.12$) are found
   using different fitting techniques, which do not account for individual position uncertainties. 
   At coarser spatial resolutions these indices get slightly steeper, and the correlation improves. 
   We find an almost linear relation and a better correlation coefficient between the 
   local $\Sigma_{\rm SFR}$ and the ISM hydrostatic pressure or the gas volume density. This 
   suggests that the stellar disk, gravitationally dominant with respect to the gaseous disk in \mm, 
   has a non-marginal role in driving the SFR. However, the tight local correlation that exists
   between the dust optical depth and the SFR sheds light on the alternative hypothesis 
   that the dust column density is a good tracer of the gas that is prone to star formation.
}
   {}

   \keywords{Galaxies: individual (M\,33) --
             Galaxies: ISM --
             Galaxies: Local Group --
	     Galaxies: spiral
            }

   \maketitle
%


\section{Introduction}

The gas-to-star conversion process is one of the most important 
ingredients for galaxy evolution. The rate at which stars form in a galaxy 
at a given epoch depends not only on the available gas reservoir but also on the 
ability of the gas to collapse and fragment. The first seminal papers 
\citep{1959ApJ...129..243S, 1963ApJ...137..758S} related the star formation rate (SFR)
to the atomic gas densities using a power law. Comparing the gas density with the 
number of young stellar objects in the solar neighborhood
\citet{1959ApJ...129..243S} derived a power law index $n=2$, i.e.
$\Sigma_{\rm SFR} \propto \Sigma_{\rm gas}^2$. Values of 
$n$ approximately 1.5 to 2.0 were further confirmed by \citet{1978A&A....68....1G}, using more 
precise data on the radial and vertical distributions of the interstellar gas and a variety of 
young stellar objects.

Papers on external galaxies often use the term Schmidt law or Kennicutt-Schmidt (KS) law 
to relate the surface density of gas to the SFR per unit surface area, 
since these quantities are the observables for external galaxies. For half a century, numerous studies have 
been done on the KS law \citep[see][ for a review]{1998ARA&A..36..189K}. For instance, 
\citet{1989ApJ...344..685K} studied how the globally averaged SFR in a galaxy traced by \Ha\ 
emission correlates with the mean atomic, 
molecular, and total (\hi +H$_2$) gas surface densities. For a sample of 15 galaxies they found a good 
correlation for the atomic and total gas densities (KS index $n_{\rm H_{tot}} = 1.3\pm0.3$), but not for 
the molecular gas density. 
One decade later, \citet{1998ApJ...498..541K} studied the relation between the total gas 
surface density and the \Ha\ SFR density averaged over the entire galaxy, 
using observations for 61 normal spiral and 36 starburst galaxies. Due to the rather large 
scatter in the KS relation, the KS index was highly dependent on the method used to fit the data. 
For the 61 normal galaxies, a least-squares fit on the SFR density yields 
$n_{\rm H_{tot}} = 1.29\pm0.18$, while a bivariate least-squares regression leads to 
$n_{\rm H_{tot}} = 2.47\pm0.39$. Likewise, these two methods to estimate the KS indices for the 
sample of 36 starburst galaxies lead to values of $n_{\rm H_{tot}} = 1.28\pm0.08$ and 
$n_{\rm H_{tot}} = 1.40\pm0.13$, respectively. The better agreement between the two fitting methods 
mainly reflects the higher gas and SFR dynamic ranges (three orders of magnitude) spanned by the sample  
of starburst galaxies with respect to the sample of normal galaxies (one order of magnitude). 
The derivation of the KS index is very sensitive to the method used to fit the data when
there is not a wide dynamical range of the variables. 

Today several questions regarding the KS law are still open. 
First, does the SFR surface density correlate better with the total (atomic plus molecular) 
gas surface density, as suggested by the globally averaged studies, or only with 
molecular surface density, since stars condense out of molecular gas? 
Second, which is the best tracer to characterize the SFR? 
Is \Ha\ a good local {\it current} SFR tracer? Does the incompleteness of the IMF
at the high mass end in regions where only small clusters form make \Ha\ an 
unreliable SFR tracer \citep{2009A&A...495..479C}? 
The far-UV radiation traces {\it recent} SFR averaged over 
a longer period of time (up to 100~Myr): how well does it correlate with the gas density 
locally? On which scale and at which wavelength does the infrared radiation trace star 
formation? How much of the evolved stellar population is contributing to dust heating? 

Third, the dependence of the KS law on the spatial scale considered is not yet clear.
It is of interest to study at which scale the KS law might break down and how the
power law index $n$ changes as we sample smaller and smaller regions.
A first step towards a spatially resolved KS law was performed using azimuthally
averaged values of SFR and gas densities 
\citep[e.g.][]{2001ApJ...555..301M,2002ApJ...569..157W,2003MNRAS.346.1215B,2007ApJS..173..524B}. 
For instance, \citet{2002ApJ...569..157W} studied a sample of 
molecule-rich spiral galaxies and found that the SFR density is more strongly correlated 
with the H$_2$ surface density than with the total gas surface density. They derived $n_{\rm H_2}=1 - 2$ 
with an average value of 1.4 using the molecular gas surface density and a radially varying 
extinction correction. For the total gas surface density the average KS index found by 
the same authors is steeper, $n_{\rm H_{tot}}\sim1.7$. 

As the resolution of the telescopes increased, the KS law has been examined 
locally at several spatial resolutions 
\citep[e.g.][]{2005ApJ...633..871C,2008AJ....136.2846B}.
However,
feedback processes linked to SF activity cast doubt on the applicability of
the KS law on very small scales. Photodissociation or photoionization radiation
from massive stars or stellar winds can locally remove the molecular hydrogen and
make \hii\ regions not spatially coincident with peaks of CO emission. This effect is
visible in \mm\ for example when comparing the \Ha\ emission map of this galaxy
with CO~$J=1-0$ line maps \citep{2003ApJS..149..343E,2004ApJ...602..723H}.
\mm\ is a galaxy with a generally low extinction, and the recent \spitzer\ maps of this galaxy 
at 8 and 24~\mi\ emission confirm that the displacement is not due to  
embedded \hii\ regions \citep{2009A&A...493..453V,2009A&A...495..479C}.

In order to address these questions,
recent space mission data in the UV \citep[{\it GALEX},][]{2007ApJS..173..185G} 
and in the IR \citep[{\it Spitzer},][]{2004ApJS..154....1W} 
have been used together with ground-based maps in the \Ha\ recombination line.
Such a multiwavelength database enables accurate diagnostics of the SFR at high angular 
resolution in nearby galaxies.
The \hi\ Nearby Galaxy Survey \citep[THINGS,][]{2008AJ....136.2563W} has 
provided \hi\ and CO maps for many nearby galaxies at an angular resolution of 500-800\,pc, 
thus offering the possibility to examine the relation between the gas surface density 
and SFR density locally, on smaller spatial scales.
\citet{2008AJ....136.2846B} find that the best correlation 
is between the H$_2$ surface density and the SFR (traced by a combination of the 
far-UV and IR surface brightness). The average KS index is: $1.0\pm 0.2$. 
The value $n_{\rm H_2} = 1$ implies that the SFR 
scales with the mass surface density of molecular gas, i.e., that the timescale and the
efficiency of SF (the fraction of gas mass converted into stars) is constant.

The THINGS result of unit slope differs from what has been 
found in other nearby galaxies by other groups using multiwavelength data. 
\citet{2007ApJS..173..572T} 
derive a KS index of 1.64 (1.87) using $\Sigma_{\rm H_2}$ 
($\Sigma_{\rm H_I+H_2}$) in molecular dominated regions of NGC\,7331.
In M\,51 a KS index of 1.56 has 
been found by \citet{2007ApJ...671..333K} for the spatially resolved 
SFR density on 500\,pc scales. The index refers to the total gas density and has been
derived by taking into account only bright \hii\ regions in the center or along the arms of
M\,51. The KS index for the H$_2$ surface density in M\,51 is slightly lower, 1.37, very similar
to that found for azimuthally-averaged quantities in the BIMA sample or previously in 
\mm\ \citep{2002ApJ...569..157W,2004ApJ...602..723H}. In M\,31 \citep{2009ApJ...695..937B} 
the best correlation at 113\,pc scales is established between the SFR density and the 
total gas density, with a KS index similar to that found for M\,51. 
The different power law indices found in the literature, $1\le n \le 3.5$, imply
that either the SF law differs from galaxy to galaxy (or from region to region such as arms
versus whole disk) or that the KS law index is very
sensitive to the method used for deriving it (spatial scale, SFR indicator, extinction corrections, 
CO to H$_2$ conversion, background subtraction, etc.). 

Due to its proximity, large angular size, and rather low inclination, the Local Group 
spiral galaxy \mm\ is a unique target to investigate the physics underlying the KS law
in a late-type galaxy. 
\mm\ has about 1/3 of its baryons in gaseous form, a small molecular fraction and a low 
dust abundance \citep{2003MNRAS.342..199C,2009A&A...493..453V}. 
Early attempts to test the KS law in \mm\ considered only the
atomic gas because of its low molecular content, given the limited spatial resolution 
and sensitivity of molecular gas observations. In the early 1970's, 
using stars and \hii\ region counts together with neutral hydrogen gas, 
\citet{1974ApJ...191..317M}, found a KS index $n_{\rm H_I} \sim 2.35 \pm 0.26$. 
\citet{1980MNRAS.190..689N} noticed that the density of \hii\ regions 
on scales of $\sim300$\,pc had a weaker dependence on the \hi\ surface density 
in the inner regions of the galaxy than in the outer ones. 
More recently, using the first unbiased census of $^{12}$CO $J=1-0$ line emission in \mm\ and the 
far-infrared emission map provided by IRAS, \citet{2004ApJ...602..723H} found a strong 
correlation between the azimuthally averaged SFR density and the 
average molecular gas surface computed for annular regions 250\,pc wide. 
For the molecular gas surface density, the KS index is $n_{\rm H_2}=1.36\pm 0.08$, 
while for the total gas surface density, it is much steeper, $n_{\rm H_{tot}}=3.3\pm0.07$. 
This steeper index is consistent with the molecular index, 
given the low molecular gas fraction in this galaxy
which seems regulated by the balance between the gas pressure (acting on the H$_2$ formation
rate) and the dissociation radiation 
\citep{1993ApJ...419L..29E,2002ApJ...569..157W,2004ApJ...602..723H,2006ApJ...650..933B}.

Another feature of \mm\ that makes it interesting for such a study is the
inconsistency of the disk stability with the ongoing SF 
\citep{1989ApJ...344..685K,2001ApJ...555..301M}. 
For many years most of the star-forming disk of \mm\ has been known to be stable according 
to the simple Toomre gravitational stability criterion \citep{1964ApJ...139.1217T} if only 
the gas surface density is considered \citep{1993ApJ...419L..29E}. 
\citet{2003MNRAS.342..199C} has shown however that the Toomre criterion 
predicts correctly the size of the unstable star forming region of the \mm\ 
disk when the stellar gravity is considered in addition to that of the gaseous disk.
Apparently the stellar disk plays an important role in driving the disk instabilities which
trigger SF. Its gravity compresses the gas and it can affect the SFR as well.
If, for example, the density, or the free fall time, of a cloud depends on the 
disk gravity perpendicular to the plane, then the SFR density might not correlate 
with the gas density alone. 
Also if SF happens only in self 
gravitating clouds, then the presence of diffuse molecular material in CO all disk 
surveys might weaken the expected correlation.

The wealth and quality of newly available data for \mm\ has drastically 
increased during the past years thanks to 
recent space missions in the UV \citep[{\it GALEX},][]{2007ApJS..173..185G} 
and in the IR \citep[{\it Spitzer},][]{2004ApJS..154....1W}. 
The high resolution and sensitivity of the recent multiwavelength database 
for \mm\ \citep{2007A&A...476.1161V,2009A&A...493..453V}
makes it an ideal target for investigating the radial and local relations between the 
various SFR tracers and gas densities. 
The spatial resolution of the CO~$J=1-0$ survey 
\citep{2003MNRAS.342..199C,2004ApJ...602..723H} of \mm\ 
is 45~arcsec, similar 
to the resolution achieved by \spitzer\ at 160~\mi\ ($\sim$40~arcsec). We can 
therefore test the
KS law at a spatial resolution of 180\,pc. 
To shed light on the physical basis of the scaling relation between SFR and gas density 
in late type spiral galaxies, we test whether a simple relation between SFR 
and gas surface density exists in \mm. We do this using several SFR tracers and
spatial resolutions applying different fitting methods.
Galaxies such as \mm, which do not have a high SFR per unit surface area and have a nearly
constant gas surface density, might not establish a tight KS relation; hence the difference between 
various fitting methods can be large.

This article is the fourth in a series dedicated to the SF in \mm, after 
\citet{2007A&A...476.1161V} \citepalias[hereafter][]{2007A&A...476.1161V}, 
\citet{2009A&A...493..453V} \citepalias[hereafter][]{2009A&A...493..453V} and 
\citet{2009A&A...495..479C} \citepalias[hereafter][]{2009A&A...495..479C}. 
It is organized as follows: Section~\ref{sec:data} presents the data and
the methodology used to derive SFRs and gas surface densities.
Section~\ref{sec:gas} compares the neutral gas distribution, atomic and molecular, 
with the \Ha\ emission line map. 
In Sect.~\ref{sec:radial}  we study the KS law considering azimuthal 
averages of different SFR and gas surface density tracers, and 
in Sect.~\ref{sec:local} we examine the local 
KS law at various spatial resolutions, using two different methods.
In Sect.~\ref{sec:local2} we investigate whether the SFR per unit area establishes
a better correlation with the gas volume density, and we outline possible biases
when using H$\alpha$ emission as SFR tracer on a local scale and the CO line
luminosities as molecular gas tracer. Our conclusions are summarized 
in Sect.~\ref{sec:conclusion}.

\section{The data sets and methodology} \label{sec:data}

Here we describe the multiwavelength data set that has been compiled
and also the photometric methods and SFR diagnostics.
These diagnostics will be used both for the azimuthally-averaged KS law 
and for the local KS law on a series of increasing spatial scales.

\subsection{Ultraviolet and \Ha\ line images}

To investigate the continuum ultraviolet (UV) emission of \mm, we use 
{\it Galaxy Evolution Explorer (GALEX)} mission \citep{2005ApJ...619L...1M} data,
in particular the data distributed by \citet{2007ApJS..173..185G}. 
A description of \galex\ observations in the far--UV (FUV, 1350--1750~\AA) 
and near--UV (NUV, 1750--2750~\AA) 
relative to \mm\ and of the data reduction and calibration procedure 
can be found in \citet{2005ApJ...619L..67T}.

To trace ionized gas, 
we adopt the narrow-line \Ha\ image of \mm\ obtained by \citet{1998PhDT........16G}. 
The reduction process, using standard IRAF\footnote{IRAF is distributed by the 
National Optical Astronomy Observatories,
which are operated by the Association of Universities for Research
in Astronomy, Inc., under cooperative agreement with the National
Science Foundation.} procedures to subtract the continuum emission, 
is described in detail in \citet{2000ApJ...541..597H}.
The total field of view of the image is $1.75 \times 1.75$~deg$^2$ ($2048 \times 2048$ 
pixels of 2\farcs028 on a side).

\subsection{Infrared images}

Dust emission can be investigated through
the mid- and FIR data of \mm\ obtained with the 
Multiband Imaging Photometer for \spitzer\ (MIPS) instrument \citep{2004ApJS..154....1W,2004ApJS..154...25R}. 
The complete set of 
MIPS (24, 70, and 160~\mi) images of \mm\ is described in \citetalias{2007A&A...476.1161V}: 
the {\it Mopex} software 
\citep{2005PASP..117.1113M} 
was used to gather and reduce the Basic Calibrated Data (BCD). We 
chose a common pixel size equal to 1\farcs2 for all images. 
The images were background subtracted, as explained in \citetalias{2007A&A...476.1161V}.
The spatial resolutions measured on the images are 
6$''$, 16$''$, and 40$''$ for MIPS 24, 70, and 160~\mi, respectively. 
The complete field-of-view observed by \spitzer\ is very large and allows us to achieve high redundancy and 
a complete picture of the star forming disk of \mm, despite its relatively large extension on the sky. 

\subsection{21-cm and millimeter data \label{sec:gasdata}}

Several data sets are available to examine the atomic and molecular gas 
distributions: these include the Westerbork Radio Synthesis Telescope (WRST) array 
data \citep[ $24''\times 48''$ spatial resolution]{1987A&AS...67..509D} and 
Arecibo single dish survey \citep[ $4'$ sp. res.]{1997ApJ...479..244C}.
For the molecular gas emission as traced by the CO~$J=1-0$ rotational line,
we can use the Berkeley Illinois Maryland Association (BIMA) array data 
\citep[ $13''$ sp. res.]{2003ApJS..149..343E} and the Five College Radio Astronomy 
Observatory (FCRAO) single dish data \citep[ $45''$ sp. res.]{2003MNRAS.342..199C,2004ApJ...602..723H}
or the map obtained by combining the two surveys as described by 
\citet{2007ApJ...661..830R}. The interferometers in general recover 
less flux than single-dish data, since they tend to filter 
the diffuse emission from structures much more extended than the primary beam 
resolution.
In fact, BIMA observations of \mm\ contain roughly half of the single-dish flux
\citep{2003ApJS..149..343E};
this implies that single-dish data are potentially more reliable to measure the total
gas column density in \mm, even though it is not clear which one establishes
a better correlation with the SFR. 
We shall use the WRST and the FCRAO data
for the atomic and molecular gas distributions respectively, which have a comparable
spatial resolution.
Following \citet{2003MNRAS.342..199C}, the CO measurements were converted to H$_2$ mass 
column densities (\msun~pc$^{-2}$), using the standard conversion factor, 
$X=2.8\times 10^{20}$~cm$^{-2}$~(K~km~s$^{-1}$)$^{-1}$.

\subsection{Star formation tracers \label{sec:sfr}}

We can use several SF tracers to test the KS law. 
Different tracers are sensitive to different timescales of SF,
and their accuracy to trace SF episodes strongly depends on the SF 
history and dust content of the galaxy under scrutiny.
\Ha\ emission traces gas ionized by massive stars in recent bursts of star
formation over timescales of 10~Myr or so.
The FUV luminosity corresponds to relatively young stellar populations ($\leq 100$~Myr),
so can be considered as complementary to \Ha\ in terms of sensitivity to short
timescales.

To convert \Ha\ and FUV emission into SFR, we first correct
both \Ha\ and FUV for extinction using the formalism  
developed by \citet{2001PASP..113.1449C}.
This empirical approach relates the extinction to the TIR and FUV luminosities.
The TIR flux (in W~m$^{-2}$~pc$^{-2}$) is the total IR flux from 3 to 1000~\mi\ 
\citep{2002ApJ...576..159D} defined as:

\begin{equation}
F{\rm (TIR)} = 10^{-14} \times [19.5\ F_\nu(24) + 3.3\ F_\nu(70) + 2.6\ F_\nu(160)]\ ,
\end{equation} 
\noindent
where $F_\nu(24)$, $F_\nu(70)$, and $F_\nu(160)$ are the MIPS flux densities in Jy~pc$^{-2}$ 
\citepalias[see also][]{2009A&A...493..453V}.
The extinction correction can then be written as:

\begin{equation} \label{eq:AV}
A_{\rm FUV} = 2.5 \times C \times \log\left(\frac{1}{1.68} \times \frac{L(\rm TIR)}{L(\rm FUV)} +
1\right) \quad .
\end{equation}

The value of $C$ is unity if considering star forming regions. Since 
we sample star forming regions and the ISM with older populations, we
adopt an average value $C=0.7$ \citepalias{2009A&A...493..453V}. For \Ha\ extinction we shall use 
$A_{\rm H\alpha} = 0.3 \times A_{\rm FUV}$.
Following the calibrations given in \citetalias{2009A&A...493..453V}, 
the SFR for the \Ha\ and FUV wavelengths are given by the following formulae:

\begin{equation} \label{eq:SFRHa}
{\rm SFR(H}\alpha)\ [{\rm M}_\odot\ {\rm yr}^{-1}] = 8.3 \times 10^{-42}\ L({\rm H}\alpha)\ [{\rm
erg~s}^{-1}] \quad ,
\end{equation}

\begin{equation} \label{eq:SFRFUV}
{\rm SFR(FUV)\ [M}_\odot\ {\rm yr}^{-1}] = 8.8 \times 10^{-44} L(\rm {FUV)\ [erg~s}^{-1}] \quad .
\end{equation}

Furthermore, \citetalias{2009A&A...493..453V} has shown that in a galaxy of low dust content such as 
\mm, multifrequency diagnostics can help to determine the SFR more accurately. 
We adopt one ``hybrid'' SFR tracer in addition to \Ha\ and FUV tracers:
the bolometric luminosity, approximated by a linear combination of the FUV and TIR 
\citep{2007ApJS..173..572T}.
This is potentially a reliable SF tracer in regions which are more extended than individual
\hii\ regions. It reflects both 
the unobscured and obscured regions of the galaxy \citepalias[see][]{2009A&A...495..479C} 
and is sensitive to a variety of star-formation timescales. 
To estimate the bolometric luminosity, 
we used the following equation, given by \citet{2007ApJS..173..572T}:

\begin{equation} \label{eq:SFRBOL}
L{\rm (bol)} =  \nu_{\rm FUV} L_{\nu, \rm obs} {\rm (FUV)} + (1 - \eta) L{\rm (TIR)} \quad .
\end{equation}

\noindent
We know that the interstellar radiation field which heats the dust 
has a contribution from the old stellar population.
Some of the IR emission may also come from dust heated by evolved stars (AGB), and thus not be
directly associated with recent SF episodes \citepalias[see][]{2009A&A...493..453V}. 
This heating might dominate for dust outside \hii\ regions. The factor $\eta$ accounts for 
this \citep{2003ApJ...586..794B,2006ApJS..164...38I}, and we shall use $\eta = 0.3$.
The bolometric SFR was calculated according to \citet{2007ApJS..173..572T}: 

\begin{equation}
{\rm SFR(bol)\ [M_\odot\ yr^{-1}]} = 4.6 \times 10^{-44} L{\rm (bol)}\ [{\rm erg~s}^{-1}] \quad .
\label{eqn:sfrbol}
\end{equation}

\noindent
This equation is based on the calibration of \citet{2006ApJS..164...38I} derived from 
Starburst~99 models \citep{1999ApJS..123....3L}, with a Salpeter IMF ($0.1-100$~M$_\odot$), 
solar metallicity, and a continuous SF.

\subsection{Radial profiles and aperture photometry}

In Sect.~\ref{sec:radial}  we shall investigate the azimuthally averaged KS law 
by correlating the azimuthally averaged values of the SFR to gas surface density.
Following \citet{2004ApJ...602..723H}, we compute $\Sigma_{\rm SFR}$ and $\Sigma_{\rm gas}$ 
as the mean values within elliptical annuli spaced by 0.24\,kpc and 
centered on the galaxy center (01$^{\rm h}$33$^{\rm m}$50\fs90, +30\degr39\arcmin35\farcs8). 
The annuli are assumed to be circular rings viewed at an inclination of 54\degr\ 
and with the line of nodes at a position angle of 22.5\degr, thus representing the 
spatial orientation of the \mm\ disk with respect to the line of sight 
\citep{2006ApJ...647L..25M}. 
We performed this analysis on molecular, atomic and total gas tracers
(from FCRAO and WRST data) and
for each of the SFR tracers as described in the previous section. 
These azimuthal averages give $\Sigma_{\rm SFR}$ and $\Sigma_{\rm gas}$
as a function of radius, the central radius of each ring.
We then correct $\Sigma_{\rm gas}$ and the radiation emitted from the
newly formed stars for the disk inclination to obtain face-on values.

In addition to analyzing azimuthal averages, we shall investigate the local KS 
law using the highest spatial resolution possible for our dataset (see Sect.~\ref{sec:local}). 
This is the 
spatial resolution of FCRAO CO~$J=1-0$ map, similar however to that 
of 21-cm and 160~\mi\ maps.
In order to compare $\Sigma_{\rm SFR}$ and $\Sigma_{\rm gas}$ locally 
we first degraded the 24, 70, 160~\mi, FUV, and \Ha\ images to the 
lowest resolution, i.e. to 45$''$, corresponding to the spatial resolution of
FCRAO molecular map. 
This corresponds to sizes of 180\,pc at the distance of \mm. This is still large
enough not to resolve out single molecular clouds: molecular clouds sizes 
in \mm\ are smaller than 100~pc \citep{2003ApJ...599..258R}.
Then, at every position observed in the FCRAO map, we performed aperture 
photometry on each of the images with a beam of 45$''$ (full width half maximum).
This gives a total of 7664 positions over the disk of \mm.
Finally, we converted the photometric results to the appropriate
surface density units to examine the local KS law.

To investigate how the KS law varies as a function of the spatial 
scale we evaluate the KS law on coarser spatial scales by averaging 
the surface densities over adjacent positions.

\section{Comparison of neutral and ionized gas distributions \label{sec:gas}}

\begin{figure}
\includegraphics[width=\columnwidth]{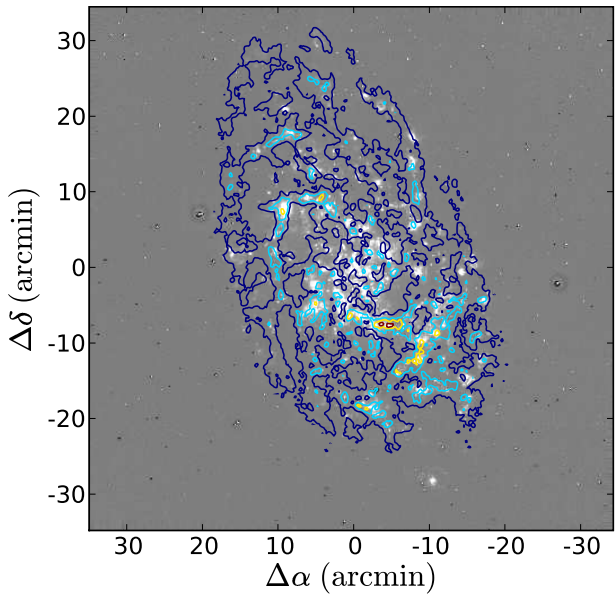}
\caption{Equal density \hi\ contours (in logarithm: 0.55, 0.85, 1.03, 1.15~M$_\odot$\,pc$^{-2}$) 
superimposed over the \Ha\ image.}
\label{fig:contourHI_Ha}
\end{figure}

Figure~\ref{fig:contourHI_Ha} shows the \hi\ contours (at logarithm levels of 
0.55, 0.85, 1.03, 1.15~M$_\odot$\,pc$^{-2}$) overplotted on the \Ha\ image. 
The \hi\ atomic gas distribution in \mm\ is very
filamentary, and the \hii\ regions lie on the high surface density \hi\ filaments.
But the bright \Ha\ emission knots in \mm\ are generally not coincident with 
the location of the neutral gas peaks.
The \hi\ contours follow the \Ha\ emission on large spatial scales, but not 
locally, on small scales. This agrees with
\citet{1972MNRAS.155..337W}, who found no correlation between the positions of the 
ten brightest \hii\ regions and the \hi\ density peaks. In the northern spiral arm, 
south of NGC\,604, 
there is relatively strong \hi\ emission, but little \Ha\ emission associated with it, as 
already noted by \citet{1980MNRAS.190..689N}. 
The \hi\ emission extends very far from the center, much beyond the optical disk, 
with a slower radial decline than that of the \Ha\ surface brightness.
The \hi\ contours even trace the northern plume 
where the \Ha\ emission is faint. 
The maxima of the \hi\ emission are in the southern half of the galaxy, 
exactly on the southern arm, and not in the geometrical center of \mm, where the gas mass
density is dominated by molecular gas, and several areas appear to be totally devoid of \hi.

\begin{figure}
\includegraphics[width=\columnwidth]{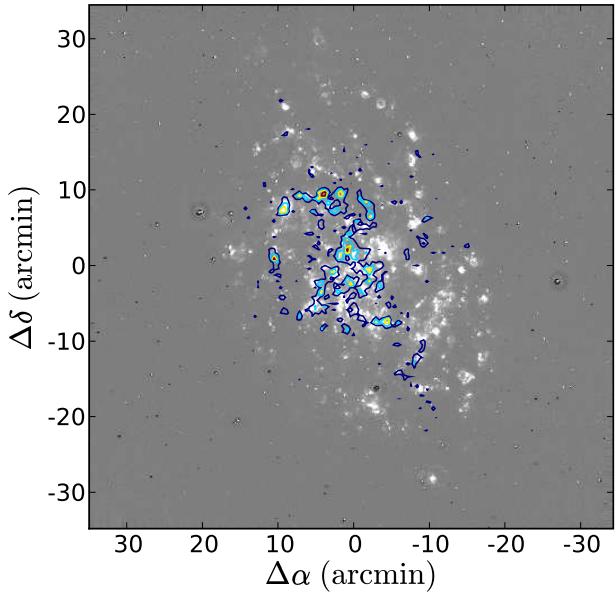}
\caption{Equal density CO contours (in logarithm: 0.90, 1.11, 1.25, 1.36~M$_\odot$\,pc$^{-2}$) 
superimposed over the \Ha\ image.}
\label{fig:contourCO_Ha}
\end{figure}

We use the FCRAO CO~$J=1-0$ emission map to trace the H$_2$ molecular gas, with
the conversion factor as given in Sect.~\ref{sec:gasdata}.
In Fig.~\ref{fig:contourCO_Ha}, 
the levels of the contours are, in logarithm, 0.90, 1.11, 1.25, 1.36~M$_\odot$\,pc$^{-2}$.
Above the detection threshold of FCRAO measurements ($3\sigma \sim 0.8$~K~km~s$^{-1}$),
molecular gas is visible only in the inner disk of the galaxy, up to about 3-4\,kpc. 
CO contours closely follow the southern spiral arm of the galaxy. In the north is
a displacement of the CO emission peaks relative to the arm as traced by H$\alpha$. 
This displacement is also evident in the BIMA data of \mm\  \citep{2003ApJS..149..343E}.
For the GMCs of BIMA, \citet{2003ApJS..149..343E} showed that there is  clustering of
GMCs and \hii\ regions out to a separation of 150\,pc. But only 67$\%$ of the detected GMCs
have their centroid position within 50\,pc of an \hii\ region.
Along the minor axis, on the western side of the galaxy, CO is absent, while there are
bright \hii\ regions visible in \Ha\ emission as well as in the IR. No CO peaks
are visible close to IC\,133, although CO emission is detected around the bright \hii\ 
region NGC\,604.

Unlike luminous spiral galaxies, the gas component of \mm\ is dominated by atomic gas 
\citep{2004ApJ...602..723H}.
The atomic-to-molecular fraction is extremely high (1 to 10) compared 
to molecule-rich galaxies \citep[e.g., 0.05 to 0.4,][]{2002ApJ...569..157W}.
Small molecular clouds have been found beyond 3-4~kpc \citep{2007A&A...473...91G}. 
The paucity of molecular gas at large radii compared to the 
atomic gas, which has no radial decline, can be explained 
as the balance between disk hydrostatic pressure and 
dissociating radiation \citep{2004ApJ...602..723H}.
Smaller clouds form in the outer disk and in some interarm regions, 
which might have a higher SF efficiency and a shorter dispersion time than elsewhere.
GMCs in \mm\ comprise less than 30\% of the molecular mass, while in our Galaxy GMCs 
contain 80\% of the molecular gas, and the fraction of molecular gas mass in GMCs 
decreases with radius.

\section{Radial Kennicutt-Schmidt law} \label{sec:radial}

We first discuss the radial variations of the KS law by analyzing
azimuthally averaged values of the SFR and gas surface density
in bins of 0.24~kpc.
To trace the SFR, 
we consider each of the three diagnostics derived in the previous section: 
extinction corrected \Ha\ (Eq.~\ref{eq:SFRHa}), 
extinction corrected FUV (Eq.~\ref{eq:SFRFUV}), 
and bolometric (Eq.~\ref{eq:SFRBOL}) luminosities. 
In Table~\ref{tab:KSlaw} we display the KS indices for all three SFR diagnostics, 
with regard to each gas component and to the total gas. 
The KS indices are rather 
similar if one considers different SFR tracers; only slightly higher KS 
indices are found when the \Ha\ emission is used to trace the SFR. We shall discuss
this effect in Sect.~\ref{sec:local2}.

The KS indices (slopes) are steeper for the total gas 
surface densities (Arecibo plus FCRAO) ($n_{\rm H_{tot}} \approx 3.1$). 
Since the atomic gas, the dominant contributor 
to the total gas surface density, has a shallow radial distribution,
we find a very high value of the KS index when we correlate the SFR with the  
atomic gas surface density alone. A tighter correlation with a shallower
slope is found between the 
azimuthally averaged SFR and the molecular gas surface density with 
$n_{\rm H_2} = 1.2\pm0.1$, from the inner disk out to 6~kpc. 
The \citet{2007arXiv0710.2102R} simulation results are in remarkable agreement 
with the radially 
averaged values of the indices found for the atomic and molecular gas in \mm.

\begin{figure*}
\includegraphics[width=1.1\textwidth]{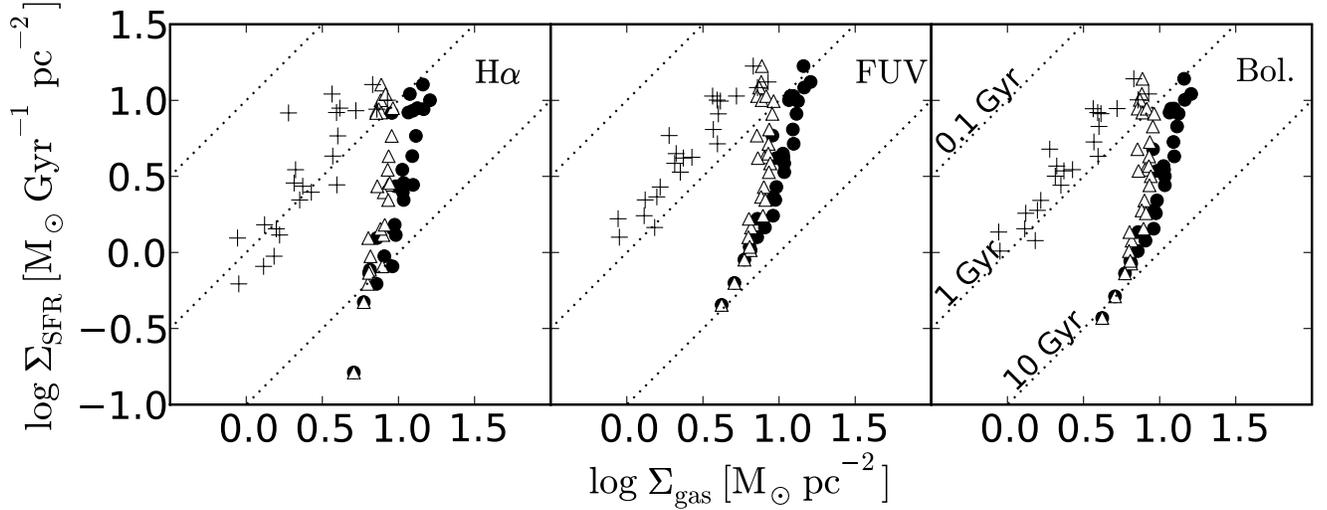}
\caption{Radial KS law, elliptically averaged over bins of 0.24~kpc. The SFR is calculated 
from three 
different tracers: extinction-corrected \Ha\ (left), extinction-corrected FUV (center), 
and bolometric (right). 
The KS law taking into account only the molecular gas surface density is displayed 
with plus symbols. The KS law relative to the atomic gas  is shown by open triangles. 
Filled circles corresponds to the KS law when the total gas surface density  is
considered. 
The dotted lines represent the gas depletion timescales (cf. labels in Gyr) 
considering a constant SF efficiency.}
\label{fig:radialKS}
\end{figure*}

The KS law is graphically presented in Fig.~\ref{fig:radialKS}, for the molecular, 
atomic, and total gas as a function of the SFR surface densities calculated from 
the three diagnostics described above.
The gas depletion timescales for a constant SF efficiency are represented by 
dotted lines. One can see that the depletion timescale for the molecular gas is rather constant 
and between 0.5 and 1~Gyr. On the other hand, the depletion timescales for the atomic and total 
gas vary widely across the disk of \mm. The fastest depletion timescales (about 0.5~Gyr) occur 
near the center of the galaxy where the SF is more pronounced; conversely, the lowest depletion 
timescales are found in the outer parts of \mm\ with values reaching roughly 10~Gyr.

\begin{table}
\begin{center}
\caption{Radially averaged KS indices ($n$) and Pearson correlation
coefficients ($r$).} \label{tab:KSlaw}
\begin{tabular}{c c c c c c c}
\hline \hline
SFR & \multicolumn{2}{c}{Molecular gas} &  \multicolumn{2}{c}{Atomic gas}
& \multicolumn{2}{c}{Total gas}\\
$n_{\rm H_2}$ & $r_{\rm H_2}$ & 
$n_{\rm H_I}$ & $r_{\rm H_I}$ &
& $n_{\rm H_{tot}}$ & $r_{\rm H_{tot}}$ \\
\hline
\Ha          & 1.3 $\pm$ 0.2 & 0.87 & 5.6 $\pm$ 0.9 & 0.76 & 3.6 $\pm$ 0.3 & 0.92 \\
FUV          & 1.1 $\pm$ 0.1 & 0.94 & 4.1 $\pm$ 0.8 & 0.73 & 2.9 $\pm$ 0.2 & 0.95 \\
Bol.         & 1.1 $\pm$ 0.1 & 0.94 & 4.1 $\pm$ 0.8 & 0.73 & 2.9 $\pm$ 0.2 & 0.95 \\
\hline
mean         & 1.2 $\pm$ 0.1 & 0.92 & 4.6 $\pm$ 0.8 & 0.74 & 3.1 $\pm$ 0.2 & 0.94 \\
\hline
\end{tabular}
\end{center}
\end{table}

Slightly higher indices than what we find using the FUV and bolometric emission
have been found by \citet{2004ApJ...602..723H} by examining radially 
averaged quantities and a SFR traced by IR emission alone (e.g., $n_{\rm H_2} \approx 1.36$).
Our KS indices are compatible with those found by \citet{2002ApJ...569..157W} if we consider the 
molecular gas alone. On the contrary, they are much higher than \citet{2002ApJ...569..157W} for 
the atomic and total gas, because their sample of galaxies is
dominated by molecular gas, while \mm\ has less than 10\% of its gas in molecular form.

\section{Local Kennicutt-Schmidt law} \label{sec:local}

In this section, 
we first examine the KS law locally at a spatial scale of 180~pc (45~arcsec),
which is the resolution of the FCRAO CO~$J=1-0$ dataset, and then at larger spatial scales.
As in the previous section, we calculate the SFR using the three different tracers and evaluate
the correlations with the atomic, molecular, and total gas surface densities. 
Extinction corrections are applied to the \Ha\ and FUV fluxes at each beam position, and
the SFR(\Ha) and SFR(FUV) are computed as described in Sect. \ref{sec:sfr}. 
At all spatial scales,
there is much dispersion in the $\Sigma_{\rm SFR} - \Sigma_{\rm gas}$ relations, 
and the dynamical range of the gas surface density is rather small. 
As a consequence, the KS indices depend on the fitting method used. We shall 
give the results from two different fitting methods, which we shall describe 
in the remainder of this section.

\subsection{Local KS law from 180 to 1440~pc: first fitting method}

\begin{figure*}
\begin{center}
\includegraphics[width=14 cm]{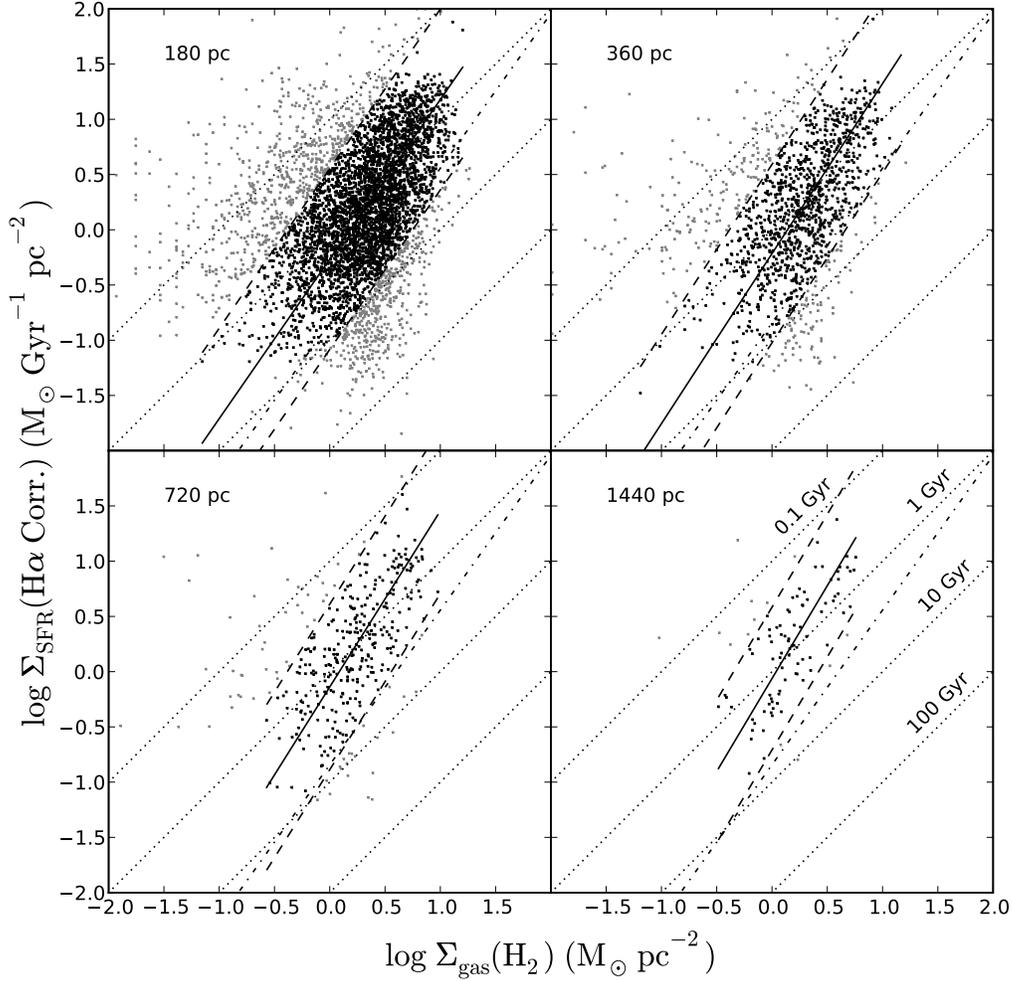}
\caption{Local KS law involving the extinction-corrected \Ha\ luminosity as SFR tracer 
and molecular gas surface density at four spatial resolutions. 
The black points (inbetween the $2\sigma$ boundaries, evidenced by the two dashed lines) 
are the ones considered for the last iteration of the first fitting method. 
The solid line is the resulting best fit considering these positions. 
The grey points were discarded during the various iterations of the fit. 
The \citet{1998ApJ...498..541K}'s index is depicted by a dash-dot line (slope of 1.4). 
The dotted lines represent the times of gas depletion (cf. labels in Gyr in the lower right panel) 
considering a constant SF efficiency.}
\label{fig:local_H2_HaCorr}
\end{center}
\end{figure*}

\begin{figure*}
\begin{center}
\includegraphics[width=14 cm]{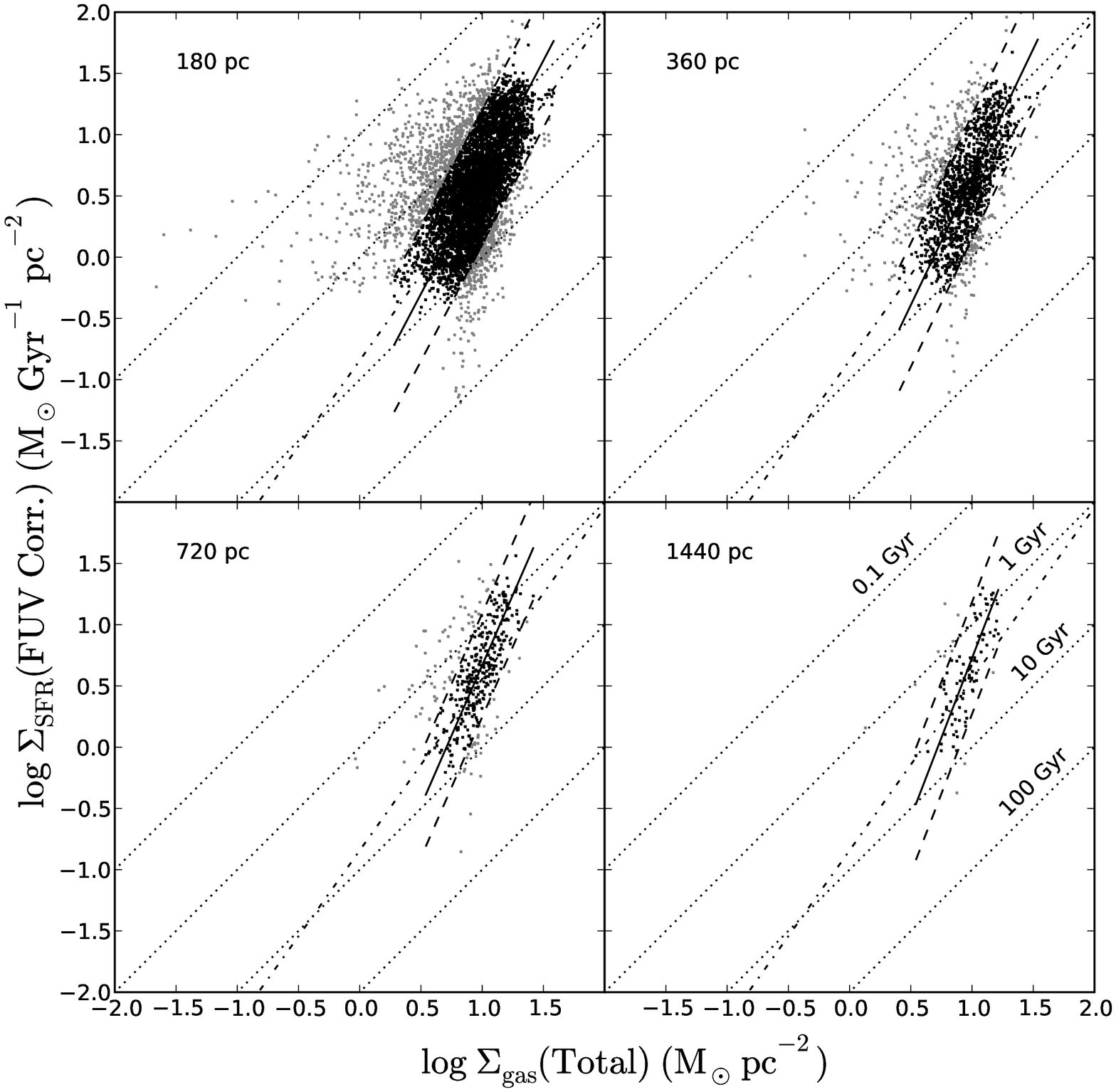}
\caption{Local KS law involving the extinction corrected FUV luminosity as SFR tracer 
and total gas surface density at four spatial resolution. 
The black points (in between the $2\sigma$ boundaries, evidenced by the two dashed lines) 
are the ones considered for the last iteration of the first fitting method. 
The solid line is the resulting best fit considering these positions. 
The grey points were discarded during the various iterations of the fit. 
The \citet{1998ApJ...498..541K}'s index is depicted by a dash-dot line (slope of 1.4). 
The dotted lines represent the times of gas depletion (cf. labels in Gyr in the lower right panel) 
considering a constant SF efficiency.}
\label{fig:local_sumTotGas_FuvCorr}
\end{center}
\end{figure*}

To estimate the indices of the KS law at a spatial resolution of 180~pc, without taking into 
account outlier positions with very high or low SFR in our map, we used a recursive fit. 
The first linear fit includes all the 6400 spatial positions defined by the central square 
($80 \times 80$ positions) of the CO map \citep{2004ApJ...602..723H}.
We estimate the dispersion, $\sigma$, of the data with respect 
to this first fit and remove all points that are lying outside $2\sigma$ before
attempting a second fit. We repeat this action recursively until no points 
lie outside the $2\sigma$ dispersion and then quote the slope of this final linear fit.
The value of the threshold has been set equal to $2\sigma$, and
changing this threshold could lead to slightly different estimates of the final KS index.
Too small a threshold will remove many points at each step and will 
end up by discarding all the data. 
On the other hand, if the threshold is too large, 
the final set of data will remain the same as the initial one, 
as no points will lie outside the boundaries. 
Ideally we want to select the value of the threshold 
which leaves as many points as possible, representing the bulk of the distribution in the 
final sample while discarding the outliers; a threshold value of $2\sigma$ seems
appropriate for this study (see Figs.~\ref{fig:local_H2_HaCorr} and \ref{fig:local_sumTotGas_FuvCorr}) 
and gives stable results.
In order to study the effect of the resolution on the KS indices, we used the same method at coarser 
spatial resolutions, by averaging the surface densities over adjacent positions: 
360, 540, 720, 900, 1080, 1260, and 1440~pc. 
In Table~\ref{tab:localKSlaw}, we show the final slopes obtained at each resolution for the 
three SFR tracers, with respect to the molecular, atomic, and total gas. 
Also in Table~\ref{tab:localKSlaw}, 
we show the Pearson coefficients obtained for the raw samples of points. Clear trends appear: 
the most significant is that the correlation between $\log \Sigma_{\rm SFR}$ and $\log \Sigma_{\rm gas}$ gets 
better and better when the resolution gets coarser. This was expected due to the stochastic effects which take 
place at small spatial scales. Although less systematic, the Pearson coefficients are in general higher when 
considering the molecular gas, with respect to the total and atomic gas. This latter shows the weakest correlation, 
independently of the SFR tracer used for the correlation.

In Figs.~\ref{fig:local_H2_HaCorr} (molecular 
gas) and \ref{fig:local_sumTotGas_FuvCorr} (total gas) we display 
the final slope obtained (heavy line) with the final samples of positions included in 
the fit, for two cases. Positions discarded (shown in grey) during the several steps of the iteration process lie outside 
the $2\sigma$ boundaries (depicted by dashed lines).
One can see that the predominant influence on the KS indices is the nature 
of the gas that is taken into account: 
the molecular gas gives values of $n_{\rm H_2}$ between 1 and 2, while atomic and total gas exhibit 
higher values: $n  \approx 2-4$. 
As in the azimuthally averaged relation, the \Ha\ SFR tracer displays higher KS indices  
than the FUV and bolometric SFR tracers; this will be discussed in the next section.

Indices get steeper on average as the resolution gets more coarse, i.e. as we average quantities
over larger areas. For the total gas density the KS indices obtained radially are even higher 
than the ones obtained for the coarser local resolution. Steeper indices for radially averaged densities are 
also found by \citet{2002ApJ...569..157W} and \citet{2007ApJS..173..572T}.
For the atomic gas the local KS indices are significantly lower, because the \hi\ 
has a shallow radial falloff, while its distribution in the disk is filamentary. Hence 
local variations of the atomic gas are more significant than radial variations. This
is not the case for the molecular gas, which declines radially with a scalelength of 2~kpc,
similar to that of the \Ha\ and FUV emission \citepalias{2009A&A...493..453V}.
The local KS indices relative to the molecular gas are similar to those found 
for azimuthal averages in radial bins. 

The number of iterations needed to reach the final slope and the data dispersion
generally decreases as a coarser spatial resolution is considered (in Table~\ref{tab:localKSlaw}
the value of the linear Pearson coefficient $r$ with the original distributions 
of points, i.e. before the first iteration, is given). This is because
stochastic effects weaken the relation between SFR and gas densities at small spatial scales. 
In the limit of \hii\ region sizes the correlation could disappear altogether,
because massive stellar winds and supernovae explosions could remove the molecular gas 
and quench subsequent SF. Using the combined BIMA-FCRAO dataset \citep{2007ApJ...661..830R}
at a spatial resolution of 30~arcsec (120~pc) around selected \hii\ regions, we find in 
fact an even weaker correlation between $\Sigma_{\rm SFR}$ and $\Sigma_{\rm H_2}$ than over 
the 180~pc spatial scale.

\begin{table*}
\begin{center}
\caption{Local KS law indices ($n$) obtained using the first fitting method for
the three SFR tracers. The number of iterations taken before the best-fit slope is 
stabilized is given in brackets. 
The Pearson coefficients ($r$) are the ones obtained with the original distributions 
of points (i.e. before the first iteration).} \label{tab:localKSlaw}
{\fontsize{6}{6}\selectfont
\begin{tabular}{ l | l l | l l | l l | l l | l l | l l | l l | l l | l l }
\hline \hline
Res. & \multicolumn{6}{c}{Molecular gas} & \multicolumn{6}{|c|}{Atomic
gas} & \multicolumn{6}{c}{Total gas} \\
(pc) & \multicolumn{2}{c}{\Ha} & \multicolumn{2}{c}{FUV} & \multicolumn{2}{c}{Bol.} 
& \multicolumn{2}{|c}{\Ha} & \multicolumn{2}{c}{FUV} & \multicolumn{2}{c|}{Bol.} 
& \multicolumn{2}{c}{\Ha} & \multicolumn{2}{c}{FUV} & \multicolumn{2}{c}{Bol.} \\
& $n$ & $r$ & $n$ & $r$ & $n$ & $r$ & $n$ & $r$ & $n$ & $r$ & $n$ & $r$ & $n$ 
& $r$ & $n$ & $r$ & $n$ & $r$ \\
\hline
180  & 1.44 (43) & 0.21 & 1.12 (24) & 0.24 & 1.12 (30) & 0.24 & 
2.53 (14) & 0.19 & 1.96 (26) & 0.23 & 1.96 (23) & 0.23 & 
2.59 (18) & 0.21 & 1.91 (19) & 0.26 & 1.90 (19) & 0.26\\
360  & 1.54 (16) & 0.26 & 1.17 (25) & 0.29 & 1.17 (28) & 0.29 & 
2.59 (12) & 0.24 & 2.10 (14) & 0.29 & 2.11 (15) & 0.29 & 
2.82  (8) & 0.27 & 2.09 (17) & 0.32 & 2.09 (18) & 0.33\\
540  & 1.64 (20) & 0.29 & 1.17  (9) & 0.33 & 1.17 (11) & 0.33 & 
2.88 (11) & 0.26 & 2.18 (10) & 0.30 & 2.24 (15) & 0.31 & 
3.39 (16) & 0.32 & 2.40 (13) & 0.38 & 2.41 (12) & 0.38\\
720  & 1.59  (7) & 0.31 & 1.16 (11) & 0.32 & 1.15  (9) & 0.32 & 
3.16 (11) & 0.31 & 2.44 (10) & 0.36 & 2.44 (13) & 0.36 & 
3.38 (10) & 0.28 & 2.29  (9) & 0.33 & 2.30 (12) & 0.33\\
900  & 1.74 (10) & 0.38 & 1.28  (6) & 0.40 & 1.28  (6) & 0.40 & 
3.29 (10) & 0.32 & 2.09  (7) & 0.39 & 2.10 (8) & 0.39 & 
3.49  (9) & 0.44 & 2.22  (6) & 0.53 & 2.24  (7) & 0.53\\
1080 & 1.81 (15) & 0.45 & 1.27 (13) & 0.55 & 1.22  (8) & 0.55 & 
3.62  (6) & 0.32 & 2.80  (7) & 0.38 & 2.85 (8) & 0.38 & 
3.98 (13) & 0.48 & 2.27  (4) & 0.56 & 2.28  (4) & 0.57\\
1260 & 1.62  (8) & 0.56 & 1.18  (4) & 0.63 & 1.17  (5) & 0.64 & 
2.77  (4) & 0.35 & 2.82  (5) & 0.40 & 2.83 (7) & 0.40 & 
4.08 (11) & 0.52 & 3.05 (11) & 0.60 & 2.88 (12) & 0.60\\
1440 & 1.68  (3) & 0.46 & 1.38  (8) & 0.51 & 1.39  (6) & 0.51 & 
2.79  (2) & 0.36 & 2.73  (3) & 0.46 & 2.74 (3) & 0.46 & 
3.18  (3) & 0.50 & 2.60  (4) & 0.61 & 2.61  (4) & 0.61\\
\hline
\end{tabular}
}
\end{center}
\end{table*}

\subsection{Relative errors: second fitting method}

One caveat of the procedure described in the previous paragraph 
is that we have considered all positions in the map to allow smoothing and
averages over larger areas. 
This implies that even positions where the molecular
gas detection was not above the noise level have been considered. 
Thus,
the reliability of the results on the smallest spatial scale relative to the
large-scale ones could be compromised because of low signal-to-noise;
as the resolution is degraded, the signal-to-noise ratio increases 
because we are averaging over larger areas.

We now fit the $\log \Sigma_{\rm SFR} - \log \Sigma_{\rm gas}$ local relation 
at the lowest spatial resolution (180~pc), taking into 
account the errors in the determination of the gas surface density
as well as in the photometry and extinction correction 
for the determination of the $\Sigma_{\rm SFR}$. 
We use the \galex\ FUV map to determine the photometric errors\footnote{It is not 
possible to determine the errors for the \Ha\ map since the map we have
is in emission measure units and it is not possible to recover the
original errors in counts.}.
In the smoothed \galex\ FUV image
the pixel-to-pixel 1-$\sigma$ noise varies between 2.9 and 8.7$\times 10^{-6}$
counts~s$^{-1}$, and we shall consider its maximum value. The 
maximum large-scale sky 1-$\sigma$ variation of sky uncertainty is 3$\times 10^{-4}$
counts~s$^{-1}$. The sky noise is much smaller than the large-scale
sky variation and can be neglected. To the photometric errors
we add the errors for extinction corrections. Following \citet{2001PASP..113.1449C}, 
we compute the extinction in the \galex\ FUV band as in Eq.~\ref{eq:AV}. 
The largest source of uncertainty in the formula are not
the photometric errors on TIR and FUV luminosities, but the
bolometric correction factor (1.68). Since the sampled areas
contain a mix of young and old stellar populations, it is not clear
how much of the TIR luminosity is due to the young stellar
population powering the FUV emission. We estimate the 
uncertainties in $A_{\rm FUV}$ to be on average 20\%. The individual 
errors on CO flux estimates can be computed from the rms noise
in individual spectra. To these we add the error on the atomic
hydrogen surface density, which is on the order of 0.7 M$_\odot$~pc$^{-2}$ (as from map
noise), if the total gas surface density is considered.

\begin{figure}
\includegraphics[width=\columnwidth]{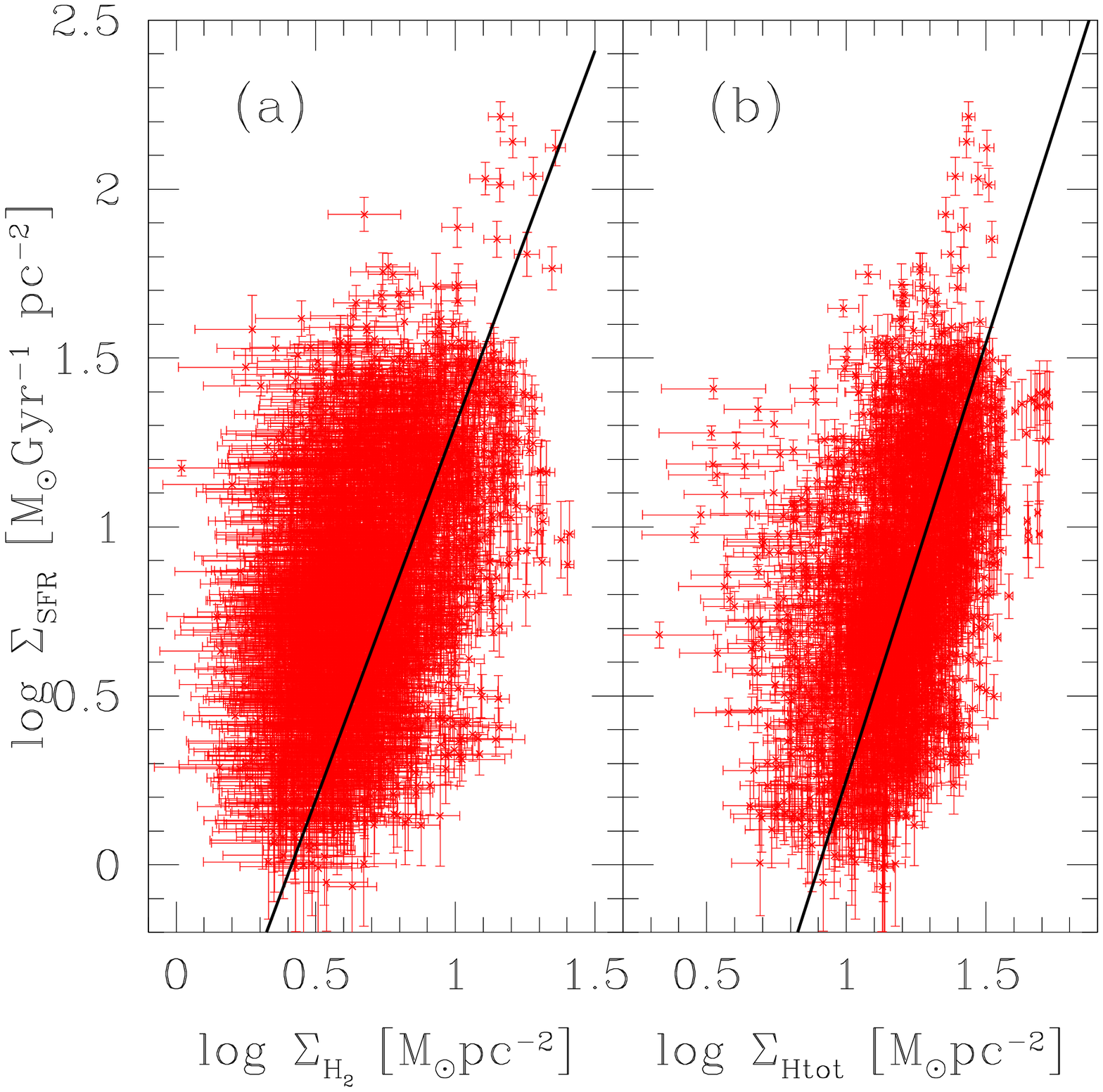}
\caption{Local KS law involving the extinction corrected FUV luminosity as SFR tracer 
and the molecular gas surface density in $(a)$, the total gas surface density in $(b)$.
The spatial resolution is 180~pc.
The best fitting linear relation leads to a KS index of $2.22\pm0.07$ in $(a)$ 
and of 2.59$\pm0.05$ in $(b)$ shown by the heavy lines.}
\label{fig:fit_err_1}
\end{figure}

We now consider all areas in \mm\ for which the CO flux
measured by the FCRAO survey is above $2\sigma$ uncertainty. 
We analyze the possible linear correlation between the $\log$ of
molecular gas surface density and the $\log$ of SFR per unit surface as traced 
by the FUV emission corrected for extinction. A linear least square method 
with errors both in the gas surface density and in the
SFR \citep[subroutine {\it fitexy} in][]{1992nrfa.book.....P} gives a slope of 
$2.22\pm0.07$. The Pearson linear correlation coefficient is 0.42.
The error on the slope is determined considering  
extinction correction error variations: 20\% on average $\pm 5\%$. 
The best-fit regression slopes and correlation coefficients relating
SFR and gas content in \mm\ for the 180~pc regions using this method 
are given in Table~\ref{tab:corre}.
For comparison, the slopes obtained with the ordinary-least-square
(OLS) method advocated by \citet{1990ApJ...364..104I} are also listed in
Table~\ref{tab:corre}. 
This method
does not take into account either uncertainties or outliers, so is less
sophisticated than either of the two methods described previously.
Indeed, the OLS slopes are shallower than either of the other two estimates;
in the case of poor correlations with many data points,
not considering properly errors or outliers can give too
much relative weight to statistically insignificant data.
In Fig.~\ref{fig:fit_err_1} we show the 
$\log \Sigma_{\rm H_{tot}} - \log \Sigma_{\rm SFR}$ relation with the relative 
best fitting linear relation (KS index = $2.59\pm0.05$). 
The correlation 
with total gas is of similar quality to that with the molecular gas alone, 
since both Pearson linear correlation coefficients are $\sim$0.43.

We now consider the correlation between the total gas 
surface density and the SFR per unit surface, including points
which have the CO brightness below the detection 
threshold. For these we shall consider only the \hi\ 
surface density. The correlation coefficient does not change, but the 
regression is shallower with a slope of $1.7\pm0.2$. 
But caution is called for in interpreting such a shallow slope as
real or claiming that there are areas in the galaxy where the
surface density decreases but the SFR does not (bi-modal 
distribution). The non-detection of molecular emission generates artificial
tails with zero slopes at low surface densities and flattens the
average slope. It is more feasible to consider positions
without a reliable CO detection only when the corresponding molecular
surface density threshold is well below the \hi\ surface density,
but there are not many positions in our data which satisfy this condition.

\begin{table}
\begin{center}
\caption{Pearson coefficients ($r$) and
slopes ($n$) derived from the ``relative errors'' method
for correlations between the two quantities 
shown in the first column. The SFR per unit area, $\Sigma_{\rm SFR}$, is traced by 
the FUV emission corrected for extinction. Only positions where the CO luminosity 
is above the 2$\sigma$ limit have been
considered at a spatial resolution of 180~pc. The surface density of molecular,
atomic and total gas are $\Sigma_{\rm H_2}$, $\Sigma_{\rm H_I}$
and $\Sigma_{\rm H_{tot}}$ respectively. $\rho$ is the volume density of the ISM  and 
$\tau_{160}$ the dust opacity. $n_1$ is the OLS slope as described in the text.
} 
\label{tab:corre}
\begin{tabular}{l c c c}
\hline \hline
&  $r$ & $n$ & $n_1$\\
\hline
$\log \Sigma_{\rm SFR} - \log \Sigma_{\rm H_2}$          & 0.42 & 2.22$\pm 0.07$  &1.46$\pm$0.34\\
$\log \Sigma_{\rm SFR} - \log \Sigma_{\rm H_I}$          & 0.24 & 1.62$\pm 0.03$  &1.11$\pm$0.15\\
$\log \Sigma_{\rm SFR} - \log \Sigma_{\rm H_{tot}}$        & 0.43  & 2.59$\pm 0.05$ &1.67$\pm$0.25\\
$\log \Sigma_{\rm SFR} - \log (\Sigma_{\rm H_2}/\rho^{-0.5}_{\rm ISM})$  & 0.62 & 1.16$\pm 0.04$ &1.00$\pm$0.01\\
$\log \Sigma_{\rm SFR} - \log \rho_{\rm ISM}$ & 0.71 & 1.07$\pm 0.02$ &0.97$\pm$ 0.01 \\
$\log \Sigma_{\rm SFR} - \log \tau_{160}$ & 0.81 & 1.13$\pm 0.02$ &1.01$\pm$ 0.02 \\
\hline
\end{tabular}
\end{center}
\end{table}

\section{Different tracers of the SFR and of the
likelihood of the gas to form stars} \label{sec:local2}

In this last section we analyze whether the SFR correlates better with other
physical quantities of the ISM, such as the hydrostatic pressure or gas volume density,
rather than with the total gas surface density. We also discuss the uncertainties
related to the use of the \Ha\ as SFR tracer, which at first seems more 
appropriate than the FUV or bolometric surface brightness in tracing the most recent star
formation. And in the last paragraph we show the tight correlation between
the SFR and the dust optical depth. We briefly discuss whether the dust optical depth
can trace the gas prone to form stars in the \mm\ disk better than 
the CO~$J=1-0$ line intensity.

\subsection{The role of the stellar disk and the dependence of the SFR on the 
gas volume density}

The widely used KS law relates the SFR per unit surface with the gas surface
density. 
The SFR depends upon the amount of molecular gas able to collapse, fragment
and form stars on a timescale $t_{\rm SF}$. Often the relevant timescale has been 
referred to as the cloud collapse or free-fall timescale, which is inversely proportional 
to the square root of the gas volume density in the cloud: 
$t_{\rm SF}\propto t_{\rm ff} \propto \rho_{\rm H_2}^{-0.5}$. 
If $\rho_{\rm H_2}$ scales with some power of the gas surface density or is constant, 
then it is conceivable that 
the SFR correlates with the surface density of the self-gravitating gas, 
$\Sigma_{\rm H_2}$, to some power. However, since the free-fall time is
much shorter than the molecular cloud formation timescale 
\citep[e.g.,][]{1994MNRAS.268..276W,2003ApJ...585..850M,2003ApJS..149..343E,2009ApJ...699..850K}, 
the relevant time for star formation could be linked to the interstellar medium
out of which the clouds form. For example \citet{2009ApJ...693.1316K} have 
recently shown using numerical simulations of disks that the ISM structure plays a primary role in 
determining the actual SFR in galactic disks. To test this hypothesis 
in \mm, we relate the star formation timescale 
not to the molecular cloud free-fall time, but to their formation timescale. 
If $t_{\rm SF}$ is the time needed to form self-gravitating molecular clouds out
of the diffuse interstellar gas, 
$t_{\rm SF}$ will depend on the density of the ISM $\rho_{\rm ISM}$, 
i.e. $t_{\rm sf} \propto \rho_{\rm ISM}^{-0.5}$ \citep{2009ApJ...693.1316K}. 
The value of $\rho_{\rm ISM}$ in the disk
scales with the hydrostatic pressure, which in galaxies with a low gaseous content such 
as \mm\ depends on the stellar disk surface
density \citep{2003MNRAS.342..199C}. An important finding in \mm\
is that both the outer SF threshold radius and the abundance of
molecules can be explained if one includes the contribution of the stellar
disk in the hydrostatic pressure equation \citep{1993ApJ...411..170E,2003MNRAS.342..199C}.

In an attempt to improve the local correlation between the gas and
the SFR in \mm, we consider the gravitational compression
of the gas due to the stellar disk. If
the hydrostatic pressure $P$ sets the gas vertical scale height $h_g$ of the disk,
the volume density $\rho_{\rm ISM}$ can be evaluated as $P/c_g^2$ or following
\citet{2003MNRAS.342..199C} as:

\begin{equation}
\rho_{\rm ISM} = {\Sigma_{\rm H_{tot}} \over 2 h_g} \quad ,
\end{equation}

\noindent
where 

\begin{equation}
h_g={c_g\over \pi G}\Bigl({\Sigma_{\rm H_{tot}}\over c_g} +
{\Sigma_s\over c_s}\Bigr)^{-1} 
\end{equation}

\noindent
and $\Sigma_s$, $c_s$ and $c_g$ are respectively the stellar surface density, the
velocity dispersion of the stars and the gas perpendicular to the disk. If 
$\rho_{\rm ISM}$ determines the SF timescale through the
cloud formation timescale (or through $t_{\rm ff}$ if 
$\rho_{\rm H_2}$ scales with $\rho_{\rm ISM}$), we expect the following relation 
to hold:

\begin{equation}
\Sigma_{\rm SFR} \propto {\Sigma_{\rm H_2} \over t_{\rm SF}} \propto
{\Sigma_{\rm H_2}\over \rho_{\rm ISM}^{-0.5}} \quad .
\end{equation}

\begin{figure}
\includegraphics[width=\columnwidth]{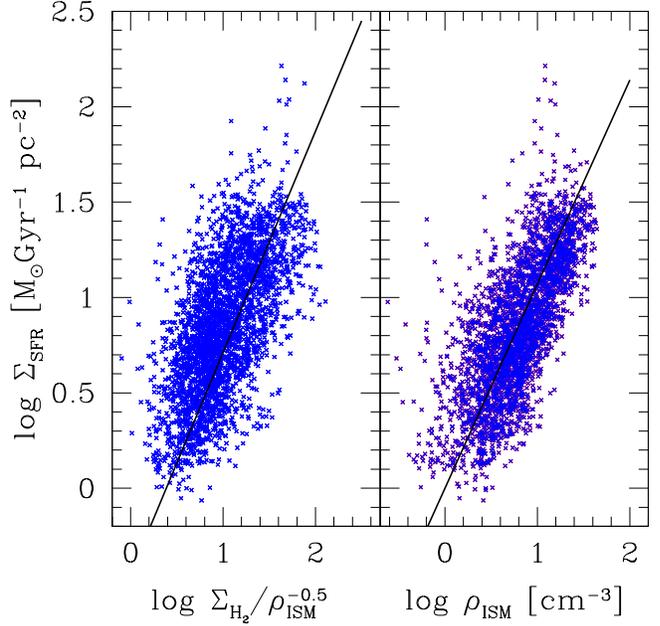}
\caption{The left-hand panel shows the linear correlation between log $\Sigma_{\rm SFR}$ and 
log $\Sigma_{\rm H_2}/\rho^{-0.5}_{\rm ISM}$ (in units of M$_\odot$/pc$^{-2}$~cm$^{-1.5}$). 
The Pearson linear correlation coefficient is 0.62, and the 
slope of the linear fit is $1.16\pm0.04$. The slope is derived taking into account uncertainties on
both axes, but they are not shown in the graph for clarity. On the
right-hand panel we show the linear correlation between $\log \Sigma_{\rm SFR}$ and $\log \rho_{\rm ISM}$
with the Pearson coefficient 0.71.}
\label{fig:fit_err_2}
\end{figure}

We shall
consider a gas velocity dispersion of 6~km~s$^{-1}$ and a ratio of 
gas-to-stellar velocity dispersion equal to 0.3 \citep{2007ApJ...669..315C}. 
By looking at  the 
correlation between $\Sigma_{\rm SFR}$ and $\Sigma_{\rm H_2}/\rho_{\rm ISM}^{-0.5}$ we find that this is
tighter than that between $\Sigma_{\rm SFR}$ and the gas
surface density alone. The Pearson linear correlation coefficient is 0.62, and the
slope of the linear fit is $1.16\pm0.04$ in the $\log - \log$ plane (see Fig.~\ref{fig:fit_err_2}),
close to linear as expected by the above equation. 
It seems clear that the gravity of the stellar disk, dominant in \mm,
plays a major role
in regulating the SFR. Variations of the gas surface density alone are insufficient 
to explain the scatter in $\Sigma_{\rm SFR}$ and imply a steep KS index as well as a loose correlation
when considering only the gas surface density.

Finally, it is noteworthy that the Pearson linear correlation 
coefficient is even higher, $r=0.71$, when considering the  
$\log \Sigma_{\rm SFR} - \log \rho_{\rm ISM}$ relation. 
The slope of the linear correlation, shown in Fig.~\ref{fig:fit_err_2}, is $1.07\pm0.02$.
The tighter relation between $\Sigma_{\rm SFR}$ and $\rho_{\rm ISM}$ (or the hydrostatic pressure 
since $\rho_{\rm ISM}\propto P$) is well known. It can be interpreted as follows: suppose that
$t_{\rm SF}$ is proportional to $\rho_{\rm ISM}^{-0.5}$ because the cloud formation timescale 
is proportional to $\rho_{\rm ISM}^{-0.5}$ and regulates the SF. 
If the molecular gas surface density (or volume density if SF is confined into a 
layer of constant scale height), which is in the form of
bound, self-gravitating units, is proportional to $\rho_{\rm ISM}^{0.57}$, 
the expected relation
is that we find $\Sigma_{\rm SFR} \propto \rho_{\rm ISM}^{1.07}$.
This would be the case for a constant star formation efficiency.  

The surface density of 
molecular gas in the form of bound, self-gravitating units can be different from $\Sigma_{\rm H_2}$
since $\Sigma_{\rm H_2}$ includes the non negligible contribution of diffuse molecular gas.
Unfortunately there are no detailed surveys of the molecular gas in \mm\ available as yet 
which would be able to constrain 
the fraction of gas in the form of bound, self-gravitating units. Only results of
GMCs surveys are available, but the molecular mass spectrum in \mm\ is steeper than $-2$ 
and hence dominated by molecular clouds of small mass \citep{2005ASSL..327..287B}.

\begin{figure*}
\hbox{
\includegraphics[width=0.5\textwidth,bb=54 174 588 711]{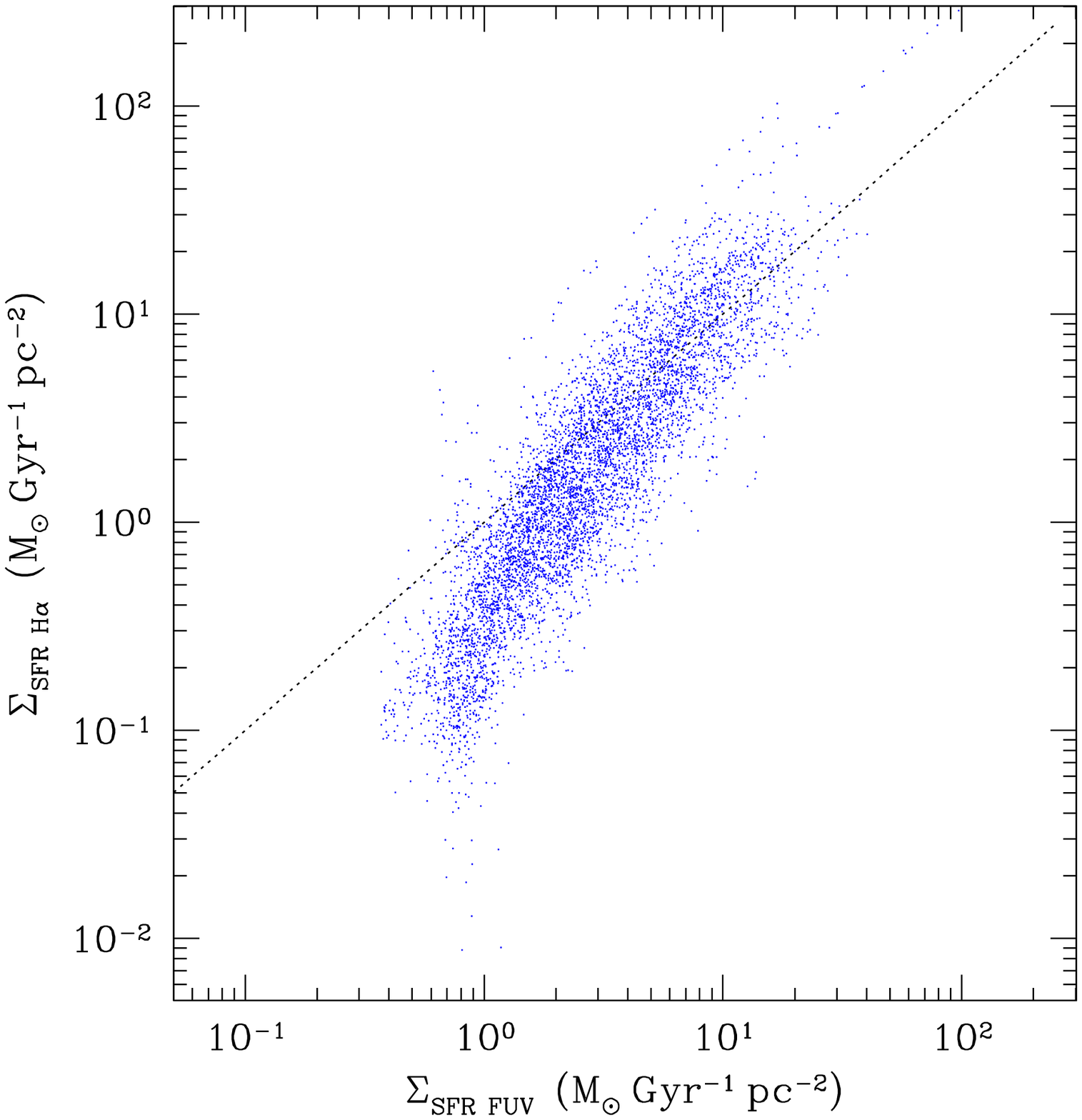}
\hspace{-0.1065\textwidth}
\includegraphics[width=0.5\textwidth,bb=54 174 588 711]{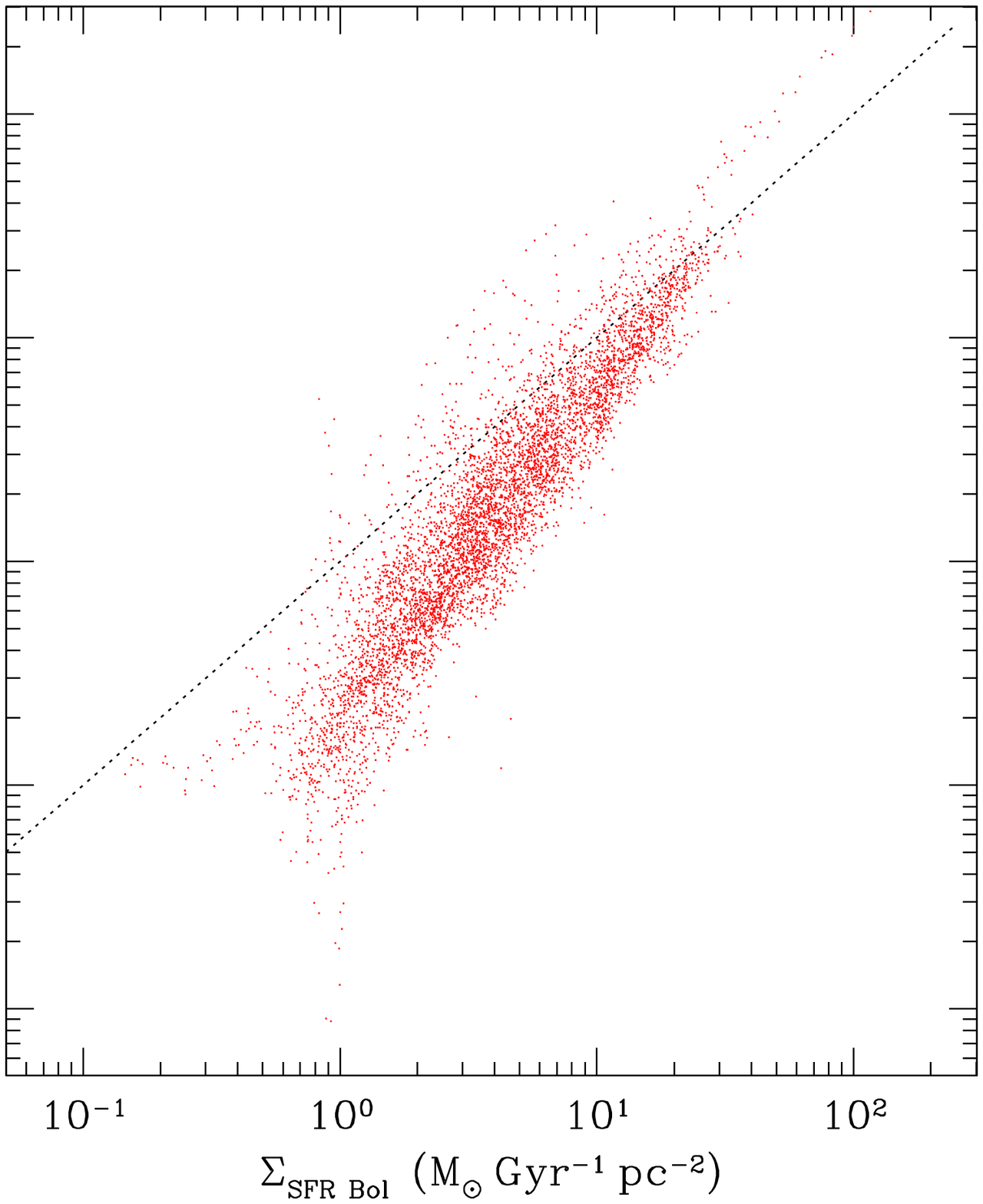}
}
\caption{$\Sigma_{\rm SFR}$ derived from \Ha\ plotted against
$\Sigma_{\rm SFR}$(FUV) (left panel) and 
$\Sigma_{\rm SFR}$(bol) (right panel). 
Only the 180~pc regions are shown.
The dotted line corresponds to equality in the two SFRs.
}
\label{fig:sfr}
\end{figure*}

\subsection{Is \Ha\ luminosity a good SFR tracer on a local scale?}

We compare the three different SFR tracers in Fig.~\ref{fig:sfr}, which
shows $\Sigma_{\rm SFR}$(\Ha) plotted against $\Sigma_{\rm SFR}$ derived
from the FUV (left panel) and bolometric luminosity (right)
(see Sect. \ref{sec:data}).
At high SFR densities, \Ha\ and FUV give quite similar values, but 
\Ha\ underpredicts SFR relative to FUV at low SFR.
SFR as inferred from bolometric luminosity is almost always larger
than that from \Ha, and at low SFR densities the effect is quite
strong.

With the \Ha\ emission as SFR indicator, the correlation of $\Sigma_{\rm SFR}$ with 
$\Sigma_{\rm gas}$ or $\rho_{ISM}$  becomes looser, and steeper indices are found.  
This is to be expected when one samples regions, as we do, with a total bolometric
luminosity below $10^{39}$~erg~s$^{-1}$, and we explain why. 
The bolometric luminosity of a single cluster or of an ensemble of 
clusters measures the mass of the clusters only when the IMF is fully populated.
Hence, when $L_{\rm bol}$ is comparable to or lower than the luminosity of the 
most massive star  in the cluster, 
the IMF cannot be fully sampled up to its high mass end.

As \citet{2009A&A...495..479C} pointed out, there are two types of
models for populating the stellar clusters when the IMF is incompletely sampled.
Either the IMF is truncated to a limiting mass which depends on the
cluster mass, or the IMF is incomplete and stochastically sampled, but maintains 
its original shape and completeness up to its high mass limit over the whole galaxy.
For both cases the L$_{\rm bol}$/L$_{\rm H\alpha}$ ratio rises when the most massive stars are lacking 
because the recombination-line luminosity,
L$_{\rm H\alpha}$ has a steeper dependence on the stellar mass than the bolometric 
luminosity does. For the first model, the truncation model,
a deviation from a simple scaling law between log~L$_{\rm bol}$ and  
log~L$_{\rm H\alpha}$ is expected when L$_{\rm bol}$ is on the order of the luminosity
of the most massive star. For 120~M$_\odot$ this is about $7\ 10^{39}$~erg~s$^{-1}$.
For the second model, the randomly sampled cluster model,
the stochastic character of the IMF implies that low luminosity 
clusters can be made either by populating the IMF up to a certain stellar mass, or  
by just one single star, or by something in between. 
The presence of outliers makes the
average L$_{\rm bol}$/L$_{\rm H\alpha}$ deviate less from its constant value at high 
luminosities (where the IMF is fully populated). In practice it is only for
L$_{\rm bol}\le 1.5\ 10^{39}$~erg~s$^{-1}$ that L$_{\rm bol}$/L$_{\rm H\alpha}$ drops 
dramatically. In this regime L$_{\rm H\alpha}$ is no longer an adequate measure 
of the cluster mass, and for this model a larger scatter is expected in the 
bolometric luminosity -- cluster mass relation. 

\begin{figure}
\includegraphics[width=\columnwidth]{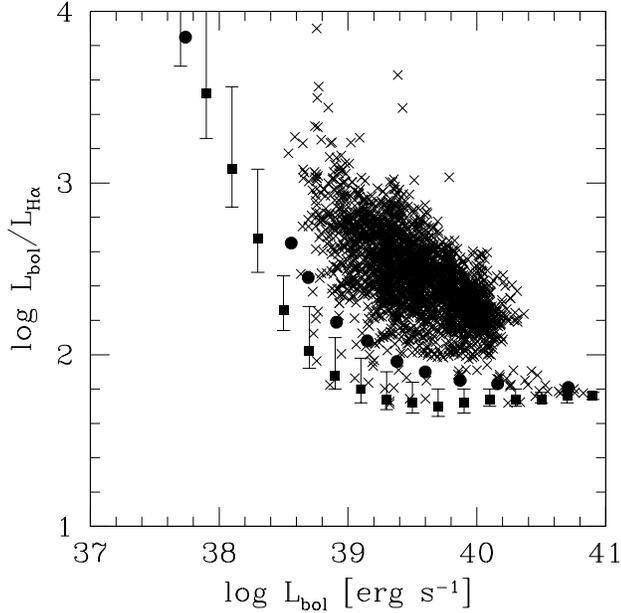}
\caption{The theoretical birthline \citepalias{2009A&A...495..479C} 
which predicts a decreasing \Ha\ flux
for decreasing luminosity of young stellar clusters due to incompleteness of the IMF 
is shown with filled squares for the randomly sampled cluster model. The
filled circles are the model prediction for clusters with a truncated IMF.
Cross symbols refer to \mm\ data, areas observed at 180~pc
resolution.}
\label{fig:birthline}
\end{figure}

Figure~\ref{fig:birthline} shows the theoretical L$_{\rm bol}$ -- L$_{\rm H\alpha}$
relation, referred to as the {\it cluster birthline}. 
The filled square symbols correspond to 
the randomly sampled cluster model, and 
filled circles indicate the prediction for clusters modelled with a truncated IMF
\citepalias[see above and][]{2009A&A...495..479C}.
Cross symbols in Fig.~\ref{fig:birthline} show the data for the 180~pc 
regions in \mm. The luminosities of these regions are clearly sufficiently low to
enter the regime of an incomplete IMF. These data cannot be used to judge which
of the two cluster models applies, because regions 180 pc wide  
may contain 
more than one cluster and not necessarily young 
\citep[see][to address this issue correctly]{2009A&A...495..479C}.

The fast decrease of the \Ha\ luminosity in regions of low
luminosity, which do not contain massive clusters, implies that using a constant 
\Ha\ luminosity $-$ SFR 
conversion factor, the inferred SFR will be much lower than the effective one. If the
inferred SFR in low density regions is lower, the index of the KS relation will 
be artificially higher. This explains the steeper 
indices we find when using \Ha\ as the SFR indicator in the local study of \mm.
We conclude that the lack of massive stars in low luminosity regions makes \Ha\ 
an unreliable tracer of the SFR. 

The birthline is the line where very young stellar clusters lie. Aging or leakage of ionizing
photons bring the clusters above the birthline. Aging decreases L$_{\rm H\alpha}$ faster 
than L$_{\rm bol}$ and moves L$_{\rm bol}$/L$_{\rm H\alpha}$ 
to higher values more or less vertically above the birthline
\citepalias[see][]{2009A&A...495..479C}. Also leakage of ionizing photons
from \hii\ regions raises the L$_{\rm bol}$/L$_{\rm H\alpha}$ values above the birthline.
All our sampled regions (\mm\ areas at 180~pc resolution) lie above the birthline (cross symbols) 
as they should, implying that we have not underestimated their bolometric luminosity. 
However, most of the SF regions we
sample are well above the birthline, and this suggests that there is a mix of
young and aged \hii\ regions in each area and/or that there is leakage of ionizing
photons.
The aging and leakage processes and the approximate formula used to estimate L$_{\rm bol}$
and extinction correction to L$_{\rm H\alpha}$ explains 
the observed scatter in the L$_{\rm bol}$ -- L$_{\rm H\alpha}$ relation. 
\mm\ is known to have a high diffuse fraction of \Ha\ emission due to the
leakage process from individual star-forming regions \citep{2000ApJ...541..597H}. 
But also in the continuum UV radiation \mm\ has high diffuse fractions
\citep{2005ApJ...619L..67T}, and if radiation escapes from the disk from these
diffuse regions or from individual star-forming sites, 
both non-ionizing and ionizing photons will contribute to this leakage.
Hence although leakage can explain a possible steepening of the 
$\Sigma_{\rm SFR}$ (from all tracers) at low gas surface densities, it does not 
explain the non-linear scaling between L$_{\rm bol}$ or L$_{\rm FUV}$
and L$_{\rm H\alpha}$.

\subsection{The correlation between the SFR and the dust optical depth}

We now use the dust opacity at 160~\mi, $\tau_{160}$
derived as in \citetalias{2009A&A...493..453V}, as an unbiased indicator of
the gas column density. In \mm\ the metallicity gradient is very shallow \citep{2007A&A...470..865M,
2008ApJ...675.1213R}, and hence it is probable that the gas-to-dust ratio does not
change much going radially outwards. The declining behavior of the dust-to-gas ratio
found in \citetalias{2009A&A...493..453V} when using the 21-cm emission line brightness
to infer the \hi\ column density and the CO~$J=1-0$ emission line brightness to infer the
H$_2$ column density, can then be  due to an underestimation of the gas surface density
at large galactocentric radii. In particular the CO-to-H$_2$ conversion factor can be
different from the value assumed in the \mm\ FCRAO survey 
($2.8\times 10^{20}$~cm$^{-2}$~(K~km~s$^{-1}$)$^{-1}$), and can vary across the
\mm\ disk according to local variations of the metal abundance and molecular clouds properties.
Many papers in the literature
\citep[e.g.][ and references therein]{2007ApJ...658.1027L,2009ApJ...702..352L} 
show that the CO-to-H$_2$ conversion factor is effectively higher in low
metallicity galaxies and can vary even in single molecular cloud complexes according to the 
cloud self-shielding conditions. As an alternative there could be opaque \hi\ gas
unaccounted for by the usual optically thin assumption for the 21-cm line emission, as found
in M\,31 by \citet{2009ApJ...695..937B}.

\begin{figure}
\includegraphics[width=\columnwidth]{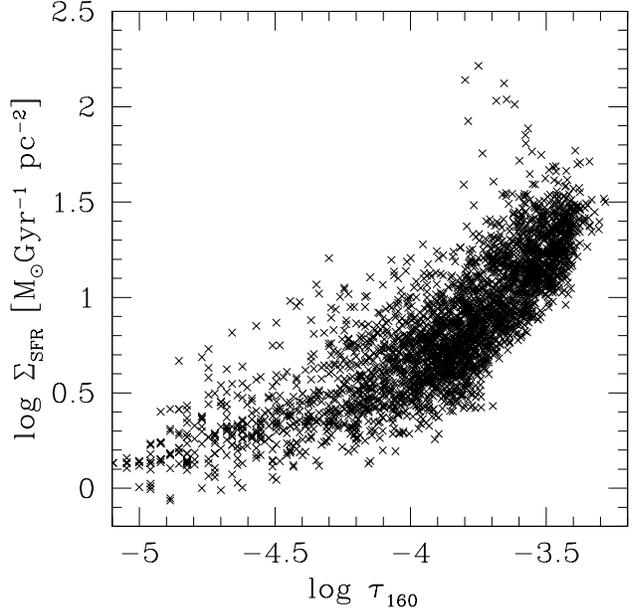}
\caption{Correlation between the SFR (FUV) surface density 
and the dust opacity at 160~\mi. Cross symbols are
data for areas in \mm\ which have a 180~pc wide diameter;
only those points are shown with the CO luminosity 
above 2$\sigma$ detection threshold.}
\label{fig:tau160}
\end{figure}

We show in Fig.~\ref{fig:tau160} the correlation between the SFR surface density
and the dust opacity.
The Pearson linear correlation coefficient is much higher ($r=0.81$) when considering
the correlation in the $\log \Sigma_{\rm SFR} - \log \tau_{160}$ plane than when using
the brightness of the 21-cm and CO~$J=1-0$ lines to derive 
$\log \Sigma_{\rm SFR} - \log \Sigma_{\rm H_{tot}}$ 
relation ($r=0.43$). As 
shown in Fig.~\ref{fig:tau160} the distribution seems bimodal;
there is an inflection in the slope around $\log \tau_{160}\approx-4$.
The average slope (given in Table~\ref{tab:corre}) 
and the correlation coefficient do not depend on whether we include positions 
where the CO luminosity is below the 2$\sigma$ limit. 
We have assumed a 20\% error on $\tau_{160}$ to determine $n$, but given
the high Pearson coefficient, the slope is not much dependent on the relative
errors.

Although the correlation
$\log \Sigma_{\rm SFR} - \log \tau_{160}$ is quite tight,
some more work is needed to check the assumptions made and 
better address the implications of this result. For example
even though the average radial variation of the metallicity is 
small ($\sim -0.4$~dex), metallicity rises in the center \citep{2009ApJ...704.1120U}, 
and at a given radius the variation in the oxygen abundance
can be almost one order of magnitude ($\sim -0.8$~dex)
\citep{2007A&A...470..843M} . Right now the relative errors on
metal abundances are large, but if future measurements confirm this
scatter, the assumption of a constant dust-to-gas ratio, implicit
in the correlation, may not be appropriate.
Also, the actual estimate of the optical depth is based on the intensity of the
emission at 160~\mi\ and on the dust temperature derived using the 70 and 
160~\mi\ emission. 
If the emission at 70~\mi\ is due to dust heated not only by the interstellar radiation 
field but also by radiation from star-forming regions,
then we would expect a higher colour temperature.
This would give a larger thermal intensity, and thus we would derive
a smaller $\tau_{160}$, anti-correlated with the SFR. High-resolution
IR observations at longer wavelengths, which will be available in the near future 
with {\it Herschel}, will allow a more accurate estimate of the 
dust optical depth. The resulting correlation coefficient involving $\log \tau_{160}$ and 
$\log \Sigma_{\rm SFR}$ could have a higher value than the one derived here.

\section{Summary and conclusions} \label{sec:conclusion}

We test the KS law in the Local Group spiral galaxy \mm, in azimuthally averaged areas
(240~pc wide) and
locally from $\sim$1.5~kpc spatial resolution down to a resolution of 180~pc. Starting from a 
multiwavelength set of observations, we used the \Ha, FUV, and bolometric (FUV+TIR)
luminosities to estimate the SFR. For gas surface density we consider the molecular, 
atomic, and total gas phases. We use extinction-corrected SFR even though extinction 
marginally affects the observed properties of the optical or 
ultraviolet emissions in \mm\  because of the rather 
low dust content of the galaxy.
The most important results are summarized below:

\begin{itemize}

\item   At every spatial scale we find that the \Ha\ KS indices are always higher than the FUV and 
   bolometric ones, and we explain this as due to the lack of \Ha\ emission in low
   luminosity regions where most of stars form in small clusters with an incomplete 
   initial mass function at their high mass end. We use the cluster birthline to support
   this, which implies a non-linear relation between the \Ha\ and bolometric luminosities. 
   The birthline also shows that most regions, even at the highest spatial
   resolution, contain a mixture of ages which on average are smaller than 10~Myr.

\item  For azimuthally averaged values, the
   depletion timescale for the molecular gas is radially constant  and the KS index is 
   1.1$\pm 0.1$ for both FUV and bolometric SFR tracer, lower than that found by 
   \citet{2004ApJ...602..723H} using IR SFR tracers alone. The depletion time for 
   the molecular gas is relatively constant, with a value of about 1~Gyr.
   The correlations with the
   molecular and total gas are tighter than with the \hi\ gas, despite the fact that 
   most of the gaseous mass of \mm\ is in \hi\ form. Denser filaments of 
   neutral gas are found where star formation takes place, but their \hi\ surface density does
   not correlate with the SFR. The KS radial index for the total gas surface density is 
   2.9$\pm 0.2$. The depletion time for the total gas density widely varies from $\sim$0.5~Gyr in the 
  center of the galaxy to about 10~Gyr in the outer parts.

\item    Locally the dispersion in the SFR-gas density relation per unit surface is high, and
results are very sensitive to the statistical method used to fit the data. The scatter in the 
KS relation increases as the spatial resolution increases. At 180~pc resolution, 
fitting the $\log \Sigma_{\rm SFR} - \log \Sigma_{\rm H_2}$ data to a straight line using 
a recursive fitting method which removes outlier points, we obtain the same the KS index 
than in the radial average analysis (1.1). There is a slight general tendency 
for the local KS laws to show higher slopes for 
coarser spatial resolutions. As for radial averages, the local KS index 
for the molecular gas is much lower than that for the total gas.

\item   For our highest resolution (180~pc) we also derive the best fitting straight line to
the data in the $\log-\log$ plane, taking into account the errors in the determination of the gas 
surface density as well as the errors in the determination of the SFR as traced by FUV emission. 
Considering only the positions 
where the CO detection is above 2$\sigma$ noise, a bivariate regression 
gives the KS index for the molecular gas of $2.22 \pm 0.07$ with a Pearson
linear correlation coefficient of 0.42. We believe that given the large scatter in the 
$\log \Sigma_{\rm SFR} - \log \Sigma_{\rm H_2}$ relation at the highest spatial resolution examined
in this paper, this method gives more robust result. This KS index is higher than the one 
we find using the recursive fit and all positions in our maps. It is also higher than what the
ordinary-least-square method gives for the same set of data (no uncertainties considered).
The bivariate regression gives a slope for the $\log \Sigma_{\rm SFR} - \log \Sigma_{\rm H_{tot}}$ 
relation only slightly higher than for the molecular gas alone: $2.64 \pm 0.07$.

\item   Given the rather poor correlations between the gas and the SFR per unit surface, 
we analyze whether the SFR correlates with other physical quantities.
A good correlation is found with the hydrostatic pressure, i.e. with the interstellar medium
volume density considering a constant sound speed. This implies that the stellar disk, 
gravitationally dominant with respect to the gaseous disk in \mm, plays a major
role in driving the SFR. The slope of the correlation is close to unity, suggesting that the
SFR per unit area (or per unit volume if the thickness of the SF disk does not vary) is
proportional to the ISM volume gas density. 
Since the correlation is tighter, its slope is less dependent on the statistical method used.

\item   There is a good correlation between the dust optical depth at 160~\mi\ and 
   the local SFR density. This can be interpreted in terms of a $\Sigma_{\rm SFR} - \Sigma_{\rm gas}$
   relation if the dust opacity is used as an
   unbiased indicator of the surface density of gas prone to star formation. 
   This might be the case for example if the gas-to-dust ratio were constant in \mm\ while the 
   CO-to-H$_2$ conversion factor were not, or if there were opaque \hi\ gas
   unaccounted for by the usual optically thin assumption for the 21-cm line emission. 
   The tightness of the
   $\log \Sigma_{\rm SFR} - \log \tau_{160}$ correlation and the shallower slope compared to the 
   $\log \Sigma_{\rm SFR} - \log \Sigma_{\rm gas}$ relation would then imply that the KS law in low 
   luminosity
   galaxies still holds, but the surface density of the gas in the process of forming
   stars has a wider dynamical range than shown by the intensity of the 21-cm and
   CO~$J=1-0$ lines. 
   Still, some caution is called for in interpreting this result
   because of some of the assumptions made in deriving it. A more detailed analysis is required
   together with a more accurate determination of the dust optical depth, which will 
   be possible in the near future thanks for example to the {\it Herschel} satellite data.
   \end{itemize}

\begin{acknowledgements}
We would like to thank Rene Walterbos for providing us the \Ha\ image of \mm, R. Kennicutt and 
G. Helou for interesting discussions, and the anonymous referee for comments that helped to improve our work. 
The work of S.~V. was supported by a INAF--Osservatorio Astrofisico di Arcetri fellowship. 
The \spitzer\ Space Telescope is operated by the Jet Propulsion Laboratory, 
California Institute of Technology, under contract with the National Aeronautics 
and Space Administration. This research has made use of the NASA/IPAC Extragalactic 
Database, which is operated by JPL/Caltech, under contract with NASA.
\end{acknowledgements}

\bibliography{astroph}

\end{document}